\documentclass{ar-1col-S2O}
\usepackage[numbers]{natbib}
\usepackage{url}
\jname{Xxxx. Xxx. Xxx. Xxx.}
\jvol{AA}
\jyear{YYYY}
\doi{10.1146/((please add article doi))}

\usepackage{graphicx}

\usepackage[normalem]{ulem}
\usepackage{xfrac}
\usepackage{dcolumn}
\usepackage{bm}
\usepackage{epsfig}
\usepackage{latexsym} 
\usepackage{amsmath,mathtools}
\usepackage{amssymb}
\usepackage{color}
\usepackage{array}
\usepackage{bbm}
\usepackage{color}
\usepackage{array}
\usepackage{cancel}
\usepackage{multirow}
\usepackage{dcolumn}
\usepackage{tabularx}
\usepackage[dvipsnames]{xcolor}
\usepackage{comment}
\usepackage{wrapfig}

\usepackage[colorlinks=true,allcolors=blue]{hyperref}%

\usepackage{titlesec}

\usepackage[normalem]{ulem} 

\newcommand{\ket}[1]{|#1\rangle}
\newcommand{\<}{\langle}
\newcommand{\e}{\varepsilon}
\newcommand{\up}{\uparrow}
\newcommand{\down}{\downarrow}
\newcommand{\Up}{\Uparrow}
\newcommand{\Down}{\Downarrow}
\renewcommand{\>}{\rangle}
\renewcommand{\(}{\left(}
\renewcommand{\)}{\right)}
\renewcommand{\[}{\left[}
\renewcommand{\]}{\right]}
\renewcommand{\v}[1]{\boldsymbol{#1}} 
\newcommand{\dslash}{d \hspace{-0.8ex}\rule[1.2ex]{0.8ex}{.1ex}}
\newcommand{\bs}[1]{\boldsymbol{#1}}
\renewcommand{\d}{\partial}
\newcommand{\del}{\nabla}
\renewcommand{\div}{\nabla\cdot}
\newcommand{\curl}{\nabla\times}
\newcommand{\eps}{\epsilon}
\newcommand{\p}{\parallel}
\newcommand{\U}{\mathcal{U}}
\newcommand{\Z}{\mathbb{Z}}
\newcommand{\T}{\mathcal{T}}
\newcommand{\tr}{\text{tr}~}
\newcommand{\N}{\mathcal{N}}
\newcommand{\C}{\mathcal{C}}
\newcommand{\E}{\mathbbm{E}}

\newcommand{\yb}{$^{171}{\rm Yb}^+$}

\newcommand{\acp}[1]{\textcolor{Aquamarine}{#1}}
\newcommand{\mff}[1]{\textcolor{Red}{#1}}
\newcommand{\gp}[1]{\textcolor{Orange}{GP: #1}}

\begin{document}

\markboth{M. Foss-Feig et al.}{Progress in Trapped-Ion Simulation}


\title{Progress in Trapped-Ion Quantum Simulation}
\author{Michael Foss-Feig,$^1$ Guido Pagano,$^2$ Andrew C. Potter,$^3$ and Norman Y. Yao$^4$
\affil{$^1$Quantinuum, 303 S. Technology Ct, Broomfield, Colorado 80021, USA}
\affil{$^2$Department of Physics and Astronomy, Rice University,
6100 Main Street, Houston, TX 77005, USA.}
\affil{$^3$Department of Physics and Astronomy, and Stewart Blusson Quantum Matter Institute, University of British Columbia, Vancouver, BC, Canada V6T 1Z1, email: acpotter@phas.ubc.ca}
\affil{$^4$Department of Physics, Harvard University, Cambridge, MA 02138, USA}
}

\begin{abstract}
Trapped ions offer long coherence times and high fidelity, programmable quantum operations, making them a promising platform for quantum simulation of condensed matter systems, quantum dynamics, and problems related to high-energy physics. 
We review selected developments in trapped-ion qubits and architectures and discuss quantum simulation applications that utilize these emerging capabilities.
This review emphasizes developments in digital (gate-based) quantum simulations that exploit trapped-ion hardware capabilities, such as flexible qubit connectivity, selective mid-circuit measurement, and classical feedback, to simulate models with long-range interactions, explore non-unitary dynamics, compress simulations of states with limited entanglement, and reduce the circuit depths required to prepare or simulate long-range entangled states.
\end{abstract}

\maketitle
{ \hypersetup{hidelinks} \tableofcontents }

\section{Introduction and overview}


Atomic ions confined in electromagnetic traps offer an unparalleled degree of isolation and precision, attributes that are essential for the realization of robust quantum information processing. The last two decades have witnessed tremendous progress in the capabilities of trapped-ion systems in the realm of both quantum computing \cite{Bruzewicz2019} and simulation \cite{Monroe2021}.  

This review surveys selected recent progress in trapped-ion quantum simulation, emphasizing both recent hardware advances and the use of digital (gate-based) approaches to quantum simulation.
Sec.~\ref{hardware_sec} reviews recent advances in qubit selection and architecture design.
Sec. \ref{sec_Hamiltonian} addresses specific developments in Hamiltonian simulation of non-equilibrium \emph{unitary} dynamics, covering selected topics such as simulation of nuclear and high-energy physics and quantum gravity, hydrodynamics, and the synthesis of new dynamical phases of matter. Finally, Sec. \ref{sec_non_unitary} reviews how hardware capabilities for performing selective measurement of qubits open opportunities for exploration of new realms of \emph{non-unitary} dynamics, enabling the creation of long-range entangled topological states with applications to quantum error correction, new forms of measurement-induced phase transitions (MIPTs), and qubit-efficient quantum simulation algorithms.


\section{Trapped-Ion Hardware Developments} \label{hardware_sec}
    
Trapped ions can be used to realize pristine qubits amenable to high-fidelity unitary and non-unitary manipulation. All the commonly used atomic species for trapped-ion quantum computing and simulation purposes share the same basic atomic structure. They are Group II or rare-earth atoms with a missing electron, which results in a hydrogen-like structure with optically closed transitions that are convenient for laser cooling and fluorescence measurements. Typically the qubit is encoded in two long-lived atomic states, either in the hyperfine manifold of the $^2S_{1/2}$ ground state \cite{Knight2003quantum} or using a ground state sublevel and an optically excited metastable electronic state\,\cite{blatt2008entangled}, such as the $^2D_{5/2}$ state    ($^2F_{7/2}$ in Yb$^+$), which is accessible with commercial lasers in heavier ions. In Table \ref{fig_Qubits}, we list the most commonly used atomic species for trapped-ion quantum computing and simulation.
Usually, the qubit states are chosen \cite{Olmschenk2007} or tuned \cite{harty2014high-fidelity} to be first-order insensitive to external magnetic fields to maximize their $T_2^*$ coherence time. 
Optical addressing, in the form of one-photon transitions in the case of optical qubits and two-photon stimulated Raman transitions in the case of hyperfine ground-state qubits, can be used for both single qubit rotations and entanglement generation operations.

\paragraph*{OMG qubits:} An interesting new development is the realization that long-lived metastable states offer magnetically insensitive pairs of states that can encode a metastable qubit \cite{Yang2022}. The presence of multiple qubit encodings in a single ion is particularly attractive and has been proposed as a new architecture for quantum computing \cite{Allcock2021}, named OMG (optical metastable ground) architecture.
In this setting, preparation, gate, and storage operations can be carried out by different qubits, and there are significant advantages in scalability because crucial operations such as sympathetic cooling and mid-circuit measurements can be carried out with reduced cross talk. Another advantage of the OMG architecture is the possibility to efficiently convert physical leakage errors into erasure errors, generally leading to higher thresholds in quantum error correcting codes \cite{Kang2023}. 

\paragraph*{Multi-level Qudits: } Along the same lines, the use of multiple states to encode quantum information (i.e.\, generalizing a qubit to a \emph{qudit}) can be realized quite naturally in ions. After the first simulation of spin-1 Hamiltonians \cite{Senko2015}, a careful theoretical study of the requirement for qudit quantum computing has been carried out in Reference~\cite{Low2020}. Single qudit manipulation has been achieved in $^{137}$Ba$^+$ with 13 states \cite{low2023control}. Trapped-ion quantum information processing has been demonstrated with up to 7 states, using the optical qubit of $^{40}$Ca$^+$ at 729 nm \cite{Ringbauer2022}. More recently, light-shift qudit-qudit gates have been demonstrated on 5 sublevels on the same ion \cite{Hrmo2023}. 

\subsection{Architecture Developments} \label{arch_sec}


Because charged particles cannot be confined in three-dimensional ($3d$) space by electrostatic potentials alone, a variety of techniques involving combinations of static and time-varying electric and magnetic fields have been developed to trap atomic ions. 
Regardless of how the confining potential is created, the basic mechanism underlying the generation of interactions between the internal states of trapped ions is largely the same. When multiple ions are laser-cooled \cite{leibfried2003quantum} in a confining potential, they equilibrate in a Wigner crystal, with their equilibrium positions and collective vibrational modes around those positions determined by the competition between the Coulomb interactions between ions and the harmonic confinement induced by the trapping potential. To generate interactions between spin (internal) degrees of freedom of two or more ions, either lasers or spatially inhomogeneous magnetic fields are used to apply forces to the ions in a spin-dependent fashion. Because the energy of two or more ions depends on their relative positions (due to their Coulomb repulsion), such forces ultimately lead to energy shifts contingent on the joint spin states of the ions, i.e.\ spin-spin interactions.  Variations of this principle underlie both the generation of spin Hamiltonians in analog simulation (see Sec.\,\ref{sec_Hamiltonian}) and the application of two- or few-qubit quantum gates.  The quality of interactions generated in this fashion can be best quantified by considering the fidelity of discrete two-qubit quantum gates that result from them.  Numerous flavors of trapped ion gates (all based on the operating principle described above) have achieved two-qubit gate fidelities close to or even slightly above $99.9\%$ (average fidelity), including M\o lmer-S\o rensen-~\cite{PhysRevLett.117.060505}, light-shift-~\cite{PhysRevLett.117.060504,clark2021high}, and magnetic field gradient driven- \cite{srinivas2021} gates. An advantage of trapped-ion qubits, in general, is that---as a result of the relatively simple atomic physics involved in the implementation of gates---the fundamental and technical limitations on two-qubit gate fidelities, such as spontaneous emission or laser phase noise, are extremely well understood. Credible pathways exist towards decreasing two-qubit gate errors by one order of magnitude, especially in the context of laser-free gates based on microwaves and/or magnetic field gradients \cite{sutherland2024}.

\subsubsection{2D ion crystals} The most popular approach to create $3d$ confinement is to use linear Paul traps \cite{paul1990electromagnetic,dehmelt1967radiofrequency} based on time-varying radiofrequency (RF) fields combined with weak static fields. However, this architecture naturally confines the ions along a single line at the null of the RF potential, where the ions can exist with minimal ``micromotion'' due to the RF driving. The use of Penning traps \cite{Brown1986Geonium} naturally extends these capabilities to a two-dimensional (2D) plane of trapped ions using a combination of static electric and Tesla-strong magnetic fields to confine particles in three dimensions. Hundreds of ions confined in 2D crystals in Penning traps have been used to simulate spin models with long-range interactions \cite{britton2012engineered}, to generate squeezing \cite{Bohnet2016} and perform quantum-enhanced sensing \cite{Gilmore2021}. However, in a single Penning trap, the ions are rotating in the lab frame at an ion-density-dependent frequency $\omega_R$ bounded by the so-called modified cyclotron frequency $\omega_c'=\frac{\omega_c}{2}-\sqrt{\frac{\omega^2_c}{4}-\frac{\omega^2_z}{2}}$ and the magnetron frequency $\omega_m=\frac{\omega_c}{2}+\sqrt{\frac{\omega^2_c}{4}-\frac{\omega^2_z}{2}}$, with $\omega_z$ being the axial frequency defined by static electrodes and $\omega_c=e B/m$ being the cyclotron frequency \cite{Thompson2016}. It is, therefore, very challenging to scale up this architecture while achieving individual unitary and non-unitary operations. 
Another complementary route to the realization of 2D ion crystals has been the use of purposely designed Paul traps to confine the ion micromotion along one plane leaving the directions orthogonal to that plane available for laser manipulations. This approach was originally proposed \cite{Richerme2016} and realized  \cite{DOnofrio2021} by using large DC confinement along the axis to ``squeeze'' the ions in a 2D plane with only radial micromotion. Another promising approach to realize 2D ion crystals with optical access to a micromotion-free direction is to rely on a three-layer slot trap as realized in Refs. \cite{Wang2020, Kiesenhofer2023}. This approach guarantees a plane where the micromotion is confined to the radial direction that is orthogonal to the laser beams, which is a desirable feature for quantum simulation and computing~ \cite{Qiao2024tunable}.
Leveraging these advances, in Reference~\cite{guo2023siteresolved}, Guo \emph{et al.} have recently demonstrated the quantum simulation of a two-dimensional, long-range, transverse field Ising model with up to 300 site-resolved ions.
The authors demonstrate the quasi-adiabatic preparation of ferromagnetic ground states and are able to directly observe spin correlations in real space, where the collective modes of the ions have been imprinted.

\subsubsection{QCCD Architectures} Because there are technical limits to the number of ions that can be effectively controlled in a single trap (RF Paul or Penning), various schemes have been explored to link together ions towards the goal of a scalable architecture with hundreds or even many thousands of ions.  The conceptually simplest approach, proposed and pioneered at the National Institute of Standard and Technology (NIST) \cite{doi:10.1142/9789814529549,wineland1998experimental,kielpinski_QCCD} and commonly referred to as the Quantum Charge Coupled Device (QCCD) architecture, is to replace the $3d$ Paul trap with electrodes printed onto a 2D substrate.  RF potentials suitable for trapping ions can still be created above the surface of the trap by applying a combination of DC and RF voltages to those electrodes \cite{chiaverini2005surface,seidelin2006microfabricated,blakestad2010}, and the ability to microfabricate intricate electrode designs in 2D enables a variety of opportunities to generate many separate trapping regions, with each one capable of holding an ion crystal.  Those separate traps can then be dynamically adjusted and repositioned, enabling the transport of ions \emph{between} traps and facilitating the generation of interactions (e.g., quantum gates) between arbitrary groups of ions. Because ions are stored in small isolated groups, the high ($\sim 99.9\%$) gate fidelities achieved in small-scale experiments can be achieved in large-scale devices \cite{H1_data}, and initialization/measurement cross-talk---a damaging and difficult source of error to suppress in many architectures---can be driven to sufficiently low levels for quantum error correction \cite{PhysRevA.104.062440}.

\begin{figure}[t]
\centering
\includegraphics[width=1\columnwidth]{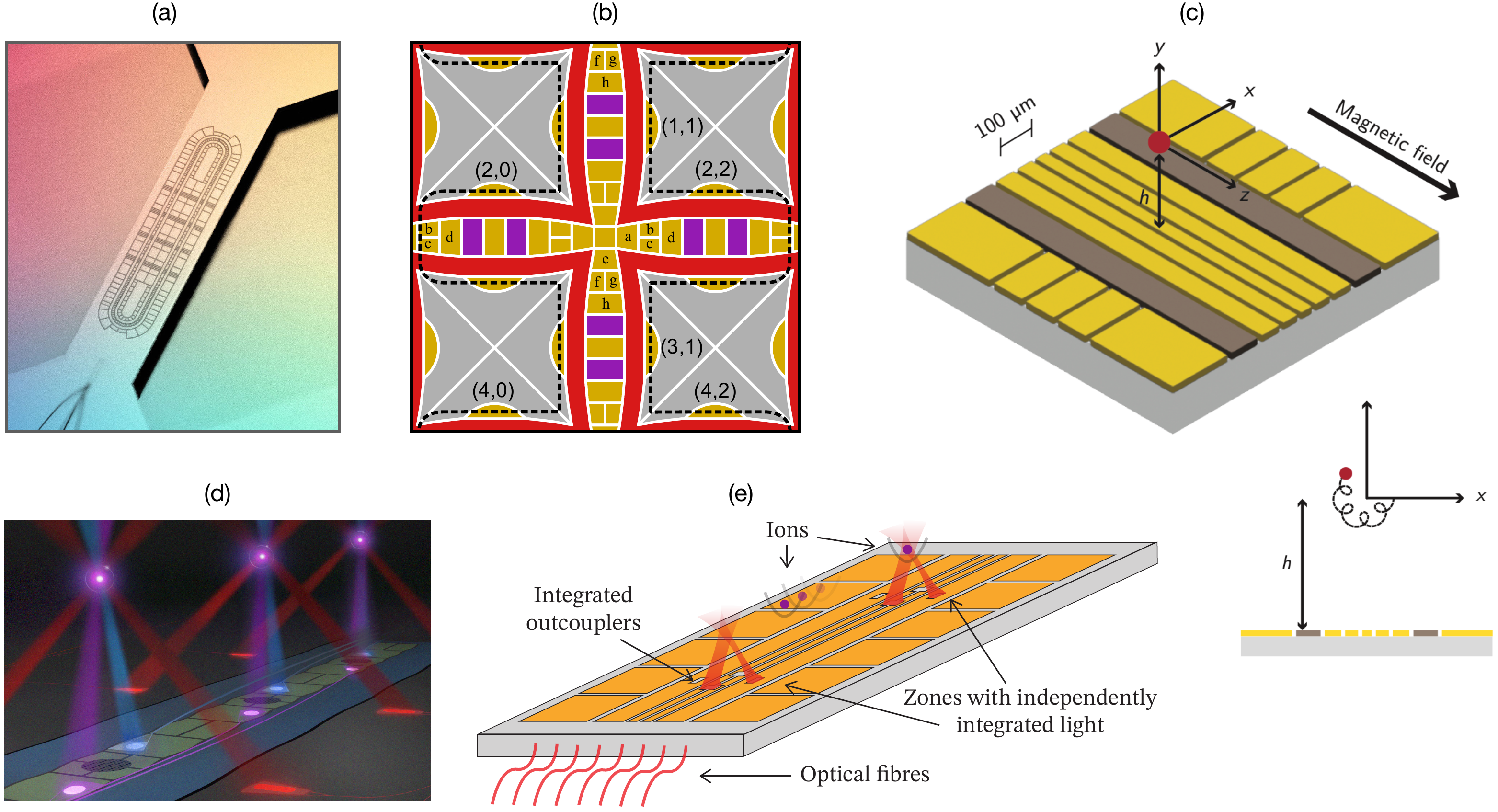}
\caption{\textbf{Recent progress in scaling the QCCD architecture:} \textbf{a)} In recent years, the QCCD architecture has been pushed to unprecedented scales, though the largest demonstrations remain fundamentally 1D. Shown above is the surface trap at the center of the H2 quantum computer (built by Quantinuum and Honeywell). \textbf{b)} Significant progress towards large-scale 2D architectures has been made, including the rearrangement of ions in a 2D grid trap using broadcast electrode signals. \textbf{c)} An exciting alternative to grid-like RF traps is the creation of micro-fabricated Penning traps in which ions are confined to (and movable within) a 2D plane. \textbf{d,e)} Scaling of the QCCD architecture benefits from the integration of light into the trap itself, routed through waveguides underneath the electrodes and focused out through the surface using grating couplers. Multiple groups have now demonstrated on-chip integration of optics and even the sequential generation of coherent operations on ions transported between multiple trap zones \cite{mordini2024multi}. Panel (a) adapted from Ref.~\cite{moses2023}. Panel (b) adapted from Ref.~\cite{delaney2024scalable}. Panel (c) adapted from Ref.~\cite{jain2024}. Panel (d) adapted from Ref.~\cite{kwon2023multi}. Panel (e) adapted from Ref.~\cite{mordini2024multi}.
}
\label{fig:daytona}
\end{figure}

Although microfabricated surface traps offer an enticing path towards storing and manipulating arrays of ions in 2D, shuttling of ions in a 2D grid poses a number of technical challenges, most crucially the ability to reliably move ions through 2D junctions with low heating. To date, the largest-scale demonstrations of the QCCD architecture have avoided the difficulties of effective junction design and transport by generating one-dimensional (1D) or quasi-1D trapping regions above a 2D surface electrode trap \cite{pino2021,moses2023}. Figure \ref{fig:daytona}(a) shows an image of the largest scale QCCD quantum computer built to date, in which ions can be transported around a 1D race-track-shaped trap. Despite the 1D geometry of the RF null (along which ion crystals can be stored with low heating), ion crystals can be rotated off the null in order to exchange the positions of the constituent ions.  Together with transport primitives that enable linear motion of crystals along the RF null and the splitting (merging) of an ion out of (into) a crystal, this capability enables arbitrary rearrangement of the qubits despite the nominally 1D geometry of the racetrack. However, it must be noted that circuits that are more highly connected than 1D geometries, e.g., the 2D circuits envisioned for fault-tolerant quantum computing based on the surface code \cite{PhysRevA.86.032324}, can only be embedded into such a 1D geometry at the cost of transport times that scale with the system size. 
Ultimately, such codes are likely to require a QCCD architecture in which qubits can be stored and moved in 2D.  A key outstanding challenge in producing a scalable 2D RF surface trap architecture has been the requirement to move multi-species crystals through junctions reliably and with low heating rates.\footnote{Multiple species enables sympathetic cooling in real-time during quantum circuits, though other approaches to this requirement include the OMG architecture described earlier.} This challenge was recently addressed in Reference \cite{PhysRevLett.130.173202} in a grid-based surface trap, and this work was extended in Reference \,\cite{delaney2024scalable} to demonstrate sorting of multi-species ion crystals using a trap-size-independent number of broadcast (i.e. each one being applied to a co-wired set of electrodes related by translational symmetry) analog voltage signals. Together, these developments suggest that large-scale 2D QCCD architectures may be achievable quite soon.

Another major technical challenge encountered in scaling QCCD architectures is the delivery of light using free-space optics. While free-space light delivery appears possible for architectures considerably larger than those in use today, the desire to build progressively larger 2D chips with ions stored close to the electrodes is ultimately at odds with addressing those ions via light shone through free space over the surface of the trap.  A promising strategy to optically address ions in larger traps while simultaneously reducing optical complexity is to deliver light through waveguides printed within the substrate supporting the electrodes and to route that light up through the surface towards the ions using grating couplers. Technological advances in waveguides, splitters, grating couplers, and other integrated optical elements have enabled remarkable progress in adding integrated optics to the QCCD toolbox. After the first demonstrations of integrated waveguides for delivery of lasers for cooling and for quantum logic operations \cite{Niffenegger2020, Mehta2020, Ivory2021}, two groups have recently demonstrated the ability to address ions in multiple independent trap zones using integrated photonics~\cite{kwon2023multi,mordini2024multi}.

Recently, a promising alternate path to achieving 2D arrays of qubits above microfabricated chips has emerged based on arrays of Penning traps  \cite{Jain2020}. By combining static electric quadrupole potentials above the chip surface with large magnetic fields, 2D arrays of ions can, in principle, be trapped without any RF signals. In this approach, there is no need to dissipate any power in the trap, alleviating one of the known challenges of scaling the QCCD architecture. Furthermore, as shown recently in Reference \cite{jain2024}, heating rates in this architecture can be $\sim$ 1 quanta/s or even lower, and transport can be performed in the whole 2D plane, relaxing another challenge of the QCCD architecture based on RF fields.

\subsubsection{Optical Interconnects} A third approach to scaling trapped-ion quantum computers is to fabricate individual, relatively small-scale traps containing tens of qubits and link them together using optical interconnects~\cite{Duan2001, Monroe2013}. In this approach, two distant modules have ions that emit two identical photons, each entangled with their respective qubit, which then interfere on a beam-splitter, creating a heralded remote entangled state among the qubits. The entangling rate can be broken down into the product \cite{Moehring2007,Hucul2015} $R_{\rm ent}=\frac{1}{2} \left(P_{\rm gen} P_{\rm coll}\right)^2 R$, where $P_{\rm gen}$ and $P_{\rm coll}$ are the photon generation and collection probabilities and $R$ is the attempt rate. All the schemes implemented so far suffer from limitations in photon collection efficiency and photon loss, leading to low entanglement generation rates compared to that for local gates. In \cite{Stephenson2020}free-space optics collection has been used to achieve $R_{\rm ent}=182\,\rm s^{-1}$ and 94\% fidelity, the best result in terms of rate and fidelity so far. One of the main bottlenecks to the remote gate speed is the photon collection efficiency. Several strategies have been attempted in this respect: for example, in-vacuum optics can lead to $P_{\rm gen} P_{\rm coll}\sim 10\%$ as shown in \cite{Carter2024}, and optical cavities can be used to both increase the collection efficiency ($P_{\rm gen}P_{\rm coll}\sim 46\%$ in \cite{Schupp2021}) and achieve more efficient fiber coupling. 


\section{Unitary Dynamics and Hamiltonian Digitization}\label{sec_Hamiltonian}
Trapped ions can natively generate pair-wise entanglement by applying a state-dependent laser force to two selected atomic states. This is usually realized by either employing a bi-chromatic drive in the so-called M\o lmer-S\o rensen scheme \cite{Molmer1999} or by applying spin-dependent light shift forces \cite{leibfried2003experimental} to the two atomic states of interest. Both schemes use the collective motional modes of the ion crystal as a quantum bus to generate ion-ion entanglement. This results in a unitary evolution operator affecting ion $i$ and $j$ that can be written as (setting $\hbar=1$ throughout) \cite{Monroe2021,schneider2012experimental}:
\begin{equation}
U_{ij}(t)=\exp{\left[
-i\zeta_i(t)\sigma_i^\alpha - i \zeta_j(t)\sigma_j^\alpha 
-i \chi_{ij}(t)\sigma^{\alpha}_i \sigma^{{\alpha}}_j  \right]},
\label{eq_ent_unitary}
\end{equation}
where the time-dependent coefficients $\zeta_i(t)$ and $\chi_{ij}(t)$ are the first-order spin-phonon coupling at site $i$ and the spin-spin interactions, respectively. Both coefficients are determined by the ratio of the strength of the spin-motional coupling and the detuning of the laser beat notes from the motional modes. The choice of the Pauli spin operator $\sigma^{\alpha=x,y,z}_i$ is controlled by the laser configuration. The M\o lmer-S\o rensen configuration gives rise to a $\sim \sigma^\phi_i \sigma^\phi_j$ interaction, with $\sigma^{\phi}_i=\sigma^x_i \cos(\phi) + \sigma^y_i \sin(\phi)$ and $\phi$ being controlled by the lasers' optical phases. In the light-shift configuration, a $\sim\sigma^z_i\sigma^z_j$ interaction term is generated. When the laser pulses are tuned up to null the first order term ($\zeta_i(t)=\zeta_j(t)=0$), and to yield $\chi_{ij}(t)=\pi/4$, the unitary realizes a maximally entangling gate (XX, YY or ZZ gate depending on the choice of $\alpha=x,y,z$, respectively), and creates a Bell state when applied to any product state in which both the two qubits lie in the plane perpendicular to the direction $\alpha$.

In the parameter regime where the spin-motion coupling is much smaller than the laser beat note detuning and the laser uniformly illuminates all the ions, the unitary evolution operator (\ref{eq_ent_unitary}) is governed by an effective long-range Ising Hamiltonian of the approximate form \cite{Monroe2021}:
\begin{equation}
    H=\sum_{i<j} \frac{J_0}{|i-j|^p} \sigma^{\alpha}_i \sigma^{{\alpha}}_j,
    \label{eq_H_long_range}
\end{equation}
where the power-law exponent $p$ can be tuned ($ 0\le p \le 3$) by changing the detuning between the laser beatnote and the center-of-mass collective mode. We note that the power-law form of the interaction is an approximation, which holds for short to intermediate scale distances~\cite{Pagano2020}; 

In addition to being able to generate two-body interactions and gates, recently, clever variations of the  M\o lmer-S\o rensen scheme have been shown to directly enable parallel gates among two pairs of qubits using more general pulse shapes \cite{Figgatt2019} and, more recently, employing two sets of orthogonal radial modes \cite{Zhu2023}. Another exciting generalization of the two-qubit gate is the realization of $N$-body entangling interactions~\cite{katz2022n,katz2023programmable}.
Rather than using state-dependent displacement forces, this approach uses state-dependent ``squeezing'' forces to generate a tunable $N$-body interaction between the ions, yielding a single-step implementation of a family of gates such as the $N$-qubit Toffoli gate (which flips a single qubit if and only if all other $N-1$
qubits are in a particular state).
Crucially, much like the original M\o lmer-S\o rensen interaction, the proposed $N$-body operation is relatively insensitive to the motional state of the ions.
Recently, in Reference~\cite{fang2023realization}, Fang \emph{et al.} experimentally demonstrated such a 5 qubit  $N$-Toffoli gate modeled on the Cirac-Zoller gate with a process fidelity estimated to be between $45$\% and $87$\%, while in Reference~\cite{katz2023demonstration}, Katz \emph{et al.} experimentally demonstrated the simulation of spin Hamiltonians comprising three- and four-body interactions.

\subsection{Simulation of Nuclear and High-Energy Physics} \label{sec_high_energy}

\begin{figure}[t]
\centering
\includegraphics[width=1\columnwidth]{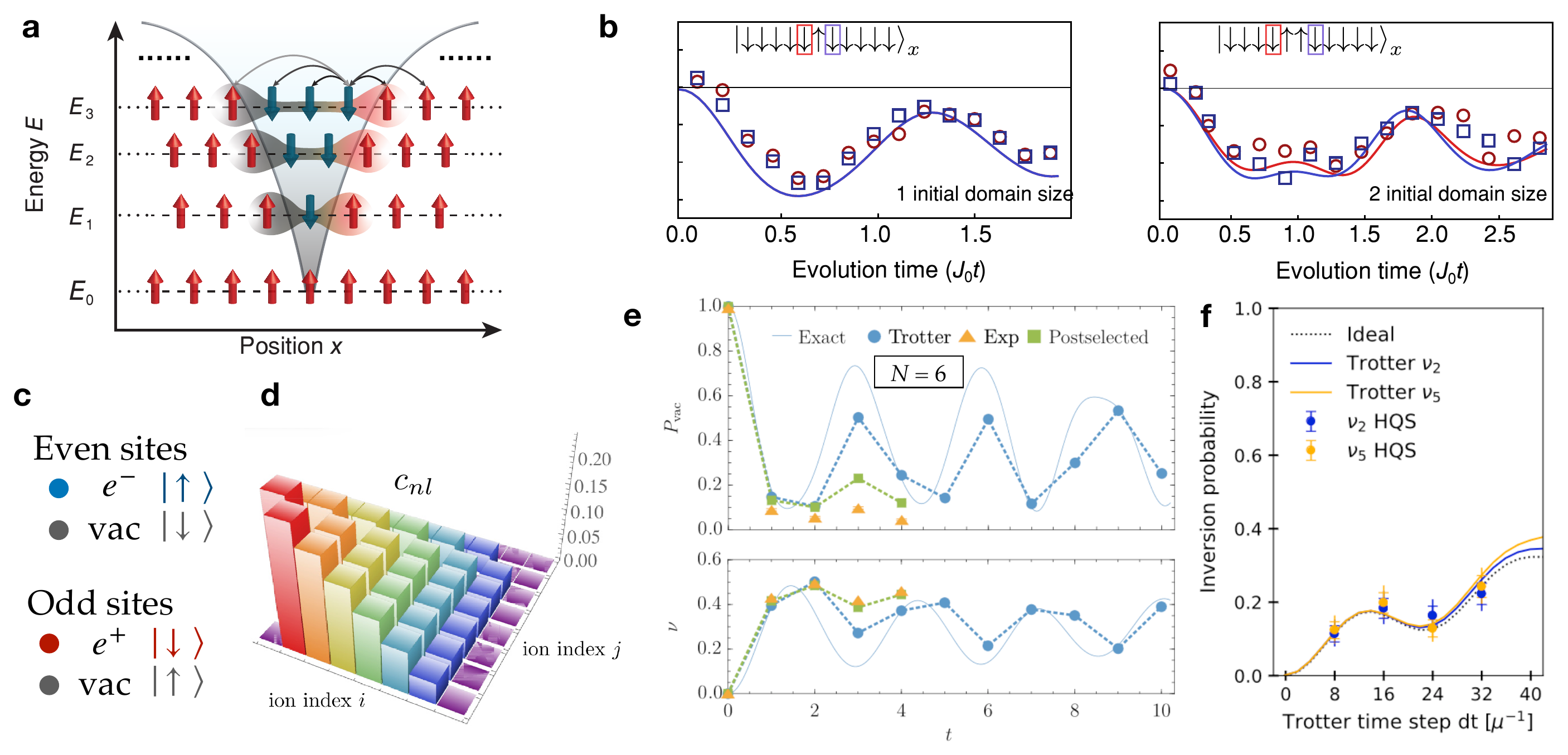}
\caption{
{\bf Trapped-ion simulations of nuclear and high-energy physics:}
{\bf a)} Magnetic domain walls in both long and short-range interacting Ising spin chains can experience an effective confining potential that increases with distance analogously to the strong nuclear force. 
{\bf b)} Magnetization oscillations ($p\sim 1.1$) starting from low energy product states to probe the first three mesons' masses, depending on the number of domain walls in the initial states. Adapted from Reference \cite{tan2021domain}.  
{\bf c)} Kogut-Susskind formulation with the Jordan-Wigner transformation allows mapping fermionic Hamiltonians onto spin systems.
{\bf d)} Spin-spin interaction matrix that encodes the Coulomb interactions in (1+1)D. Each qubit interacts with equal strength on one side and with linear decay on the other side. Adapted from \cite{Davoudi2020towards}.
{\bf e)} Results from \cite{Nguyen2022digital} showing the evolution of the particle density $\nu$ and the probability of being in the vacuum state $P_{vac}(t)=|\langle \psi_0|e^{-i H t} |\psi_0\rangle|^2$, with $\ket{\psi}$ being the initial state, for $N=6$ lattice sites.
{\bf f)} Inversion probability for the simulation of $N=8$ neutrinos starting from initial state $\ket{\Psi_0}=\ket{00001111}$. Neutrinos related by the exchange symmetry $\nu_k\leftrightarrow \nu_{N-1-k}$ are shown in the same panel. Adapted from \cite{Amitrano2023}.
}
\label{fig_high_en}
\end{figure}


In addition to simulating lattice spin models relevant to condensed matter systems, the tunable and long-range interactions of trapped-ion systems also enable simulation of lattice-discretizations of strongly-coupled quantum field theories of interest in high-energy particle and nuclear physics applications~\cite{preskill2018simulating, Banuls2020, Bauer2023, Bauer2023Quantum}. 
Early experiments in the Innsbruck group simulated the Dirac equation in 1+1D, coupling one qubit to the momentum of a single collective mode and observing paradigmatic relativistic phenomena such as Zitterbewegung \cite{Gerritsma2010dirac}. 

As shown in Eqs. (\ref{eq_ent_unitary}) and (\ref{eq_H_long_range}), trapped ions can natively realize Ising-type interactions, and transverse-field Ising models are connected to lattice gauge theories by duality transformations \cite{Wegner1971, Balian1975}. Via this duality, trapped ions can naturally realize phenomena such as confinement of mesonic excitations \cite{kormos2017real-time} and string breaking \cite{Verdel2020real}. In this setting, the mesonic excitations are mapped to domain walls whose separation is energetically suppressed by either a longitudinal field in the short-range Ising model \cite{Mazza2019Suppression} or by frustration in the long-range interaction case \cite{liu2019confined,lerose2019quasilocalized}, causing a ladder of discrete meson states to appear in the low-energy spectrum of the system \cite{James2019Nonthermal}. These phenomena have been observed experimentally for an ion chain with long-range interactions in\,\cite{tan2021domain}, where the mass scaling of the low energy part of the spectrum and the first few mesonic bound states have been measured (see Fig.\,\ref{fig_high_en}[a,b]). 

The Kogut-Susskind staggered formulation of lattice gauge theories~\cite{Kogut1975} offers a convenient mapping of fermions to spin models. Such a formulation has allowed trapped-ion simulators to address real-time dynamics of lattice gauge theories.
This discretized Hamiltonian formulation maps the presence of a particle(anti-particle) onto an occupied even (unoccupied odd) lattice site. A Jordan-Wigner transformation can be used to map the fermionic Hamiltonian to a spin Hamiltonian, where the spin state $\ket{\!\up}_n(\ket{\!\down}_n)$ corresponds to the presence(absence) of an electron on site $n\in \rm even$, and the absence(presence) of a positron on $n\in \rm odd$ (see Fig. \ref{fig_high_en}c).
In the special case of quantum electrodynamics (QED) in 1+1D (the Schwinger model), the gauge field operators can be conveniently eliminated \cite{Muschik_2017}, at the expense of introducing long-range interactions described by the spin Hamiltonian:
\begin{equation}
H = w \sum_{n=1}^{N-1} [\sigma^+_n \sigma^-_{n+1} + \sigma^-_n \sigma^+_{n+1} ] + \frac{m}{2} \sum_{n=1}^N (-1)^n \sigma^z_n +  \frac{J}{2} \sum_{n=1}^{N-2}\sum_{l=n+1}^{N-1} c_{nl} \sigma^z_n \sigma^z_l,
\label{eq_schw}
\end{equation}
where $\sigma^\pm_n$ are the raising/lowering spin operators on site $n$, while $w$, $m$, and $J$ set the energy scales of the hopping, the masses of the fermions, and gauge field interactions, respectively. The coefficients $c_{nl}$ (Fig. \ref{fig_high_en}c)  are set by translational invariance and the gauge constraints given by the Gauss law, expressed in terms of the electric field operators $E_{n}=E_{n-1} + \frac{1}{2}[\sigma^z_n+(-1)^n]$. This formulation has the advantage of enforcing the gauge constraints, as the Hamiltonian commutes with the Gauss law operators $G_n=E_n-E_{n-1}-\sigma^z_n + \frac{1}{2}(1-(-1)^n)$. The Trotterization of Hamiltonian (\ref{eq_schw}) with $N=4$ sites has been realized in \cite{Martinez2016real} using XX and YY gates for the nearest-neighbor hopping, individual rotation for the mass terms and employing elaborate local shelving and de-shelving operations combined with all-to-all Molmer-Sorensen interactions to realize the gauge field term in the Hamiltonian (\ref{eq_schw}). This set of operations enabled the simulation of the real-time dynamics of the model, observing both the particle number and the vacuum persistence probability (i.e., the probability of returning to the initial state, also known as Loschmidt echo). The same Hamiltonian has also been simulated in a digital setting in \cite{Nguyen2022digital} with more flexible pair-wise Molmer-Sorensen operations that allowed the measurement of the same observables for longer times and with up to $N=6$ sites (see Fig. \ref{fig_high_en}e). In both \cite{Martinez2016real} and \cite{Nguyen2022digital}, the initial state was the vacuum of the theory with dominant electric-field interactions, which is mapped to an antiferromagnetic product state.

The ground state of the Schwinger Hamiltonian (\ref{eq_schw}) has also been simulated experimentally with up to $N=20$ spins in \cite{Kokail2019} using a variational approach. The addition of a topological term $\sim \theta\sum_n E_n$ to the Hamiltonian (\ref{eq_schw}) makes the Schwinger model a prototype model to study charge conjugation parity (CP) violation in Quantum Chromodynamics (QCD). A sufficiently strong quench in the $\theta$ parameter has been predicted \cite{Zache2019} to lead to a Dynamical Quantum Phase Transition \cite{jurcevic2017direct,zhang2017observation,heyl2018dynamical} which was realized experimentally in \cite{Mueller2023quantum} using IonQ quantum hardware.

Thus far, all the realizations of the Schwinger model have been implemented with digital protocols. Analog Hamiltonian simulations using trapped ions have been proposed theoretically via the direct realization of pure three-body interactions \cite{Hauke2013, Andrade2022}, using multiple motional modes \cite{Davoudi2020towards} or using hybrid digital-analog approaches using bosonic excitations to encode directly the gauge-link and electric-field operators \cite{Davoudi2021towards}. 
All these approaches would realize Abelian field theories, while analog schemes to realize non-Abelian theories remain an open challenge. In particular, since gauge fields formally live in infinite-dimensional Hilbert spaces, their efficient encoding onto quantum simulators made of qubits is an outstanding challenge. One interesting approach involves the use of qudits: in \cite{meth2023simulating}, qudits with up to 5 states are used to encode the gauge fields of 2D QED. Here, the coupling between a matter site, encoded in a qubit, and the gauge field is realized as a two-body qubit-qudit interaction. Along the same lines, it has been recently proposed in \cite{calajò2024digital} that qudit-qudit gates can be used for digital simulation of non-Abelian gauge theories.

Another prominent research direction aims to simulate, using trapped-ion systems, scattering amplitudes, which is a pivotal quantity in particle physics that has been addressed in a number of theoretical proposals ~\cite{Pichler2016, Surace_2021, Milsted2022, bennewitz2024simulating}. A first proof-of-principle study was conducted in \cite{Gustafson2021} using a trapped-ion quantum computer to extract the phase shift of a scattering process measuring the time delay between the free and interacting short-range Ising models (see Fig. \ref{fig_high_en}b). Recently, Quantinuum quantum hardware has been used to prepare ansatz of interacting mesonic wave packets using variational quantum algorithms \cite{davoudi2024scattering} and to simulate a $beta$-decay of a baryon on a single lattice site using 20 qubits \cite{Farrell2023}.

Trapped-ion quantum computers can also be used to address models related to collective flavor oscillations of neutrinos caused by forward neutrino-neutrino scattering in high-density regimes such as \cite{Pantaleone1992, Cervia2019}. It has been demonstrated that a system with only two flavors can be mapped to an all-to-all Heisenberg-like spin Hamiltonian \cite{Pehlivan2011} of the form:
\begin{equation}
\label{eq_H_nu}
H=\sum_{p} \vec{B}_p \cdot \vec{\sigma}_p + \mu \sum_{ p,q}(1-\cos({\bf \theta}_{p,q}))\vec{\sigma}_{ p}\cdot\vec{\sigma}_{ q}.
\vspace{-0.2cm}
\end{equation}
The one-body free term is responsible for neutrino vacuum oscillation with the effective magnetic fields expressed in the flavor and mass basis as $\vec{B}_p=\frac{\delta m^2}{2p}\,(\sin(2\theta),0,-\cos(2\theta))_{\rm flavor}=\,\frac{\delta m^2}{2p}(0,0,-1)_{\rm mass}$ where $\delta m$ and $\theta$ are the mass difference and the two-flavor mixing angle, respectively. The interaction term takes into account the forward scattering processes, where $\mu=G_F/2\sqrt{2}V$ is given by the Fermi constant $G_F$ and $\theta_{p,q}$ the scattering angle between the momentum $p$ and $q$. Recently, this model has been realized on Quantinuum quantum hardware \cite{Amitrano2023} using Trotterized dynamics simulating the population inversion of $N=4$ and $N=8$ neutrinos (see Fig. \ref{fig_high_en}f)  starting from the initial product states $\ket{0011}=\ket{\nu_e\nu_e\nu_x \nu_x}$ and $\ket{00001111}=\ket{\nu_e\nu_e\nu_e\nu_e\nu_x\nu_x \nu_x \nu_x}$, respectively. 

Finally, we note that trapped ions processors have been used recently to implement scrambling unitaries \cite{Landsman2019} and chaotic spin models such as the Bertini-Kos-Prosen model (BKP) to investigate teleportation protocols inspired by quantum gravity \cite{Shapoval2023towardsquantum}.

\subsection{Quantum metrology and hydrodynamics with long-range interactions}
%
\begin{figure}[t]
\centering
\includegraphics[width=1.0\columnwidth]{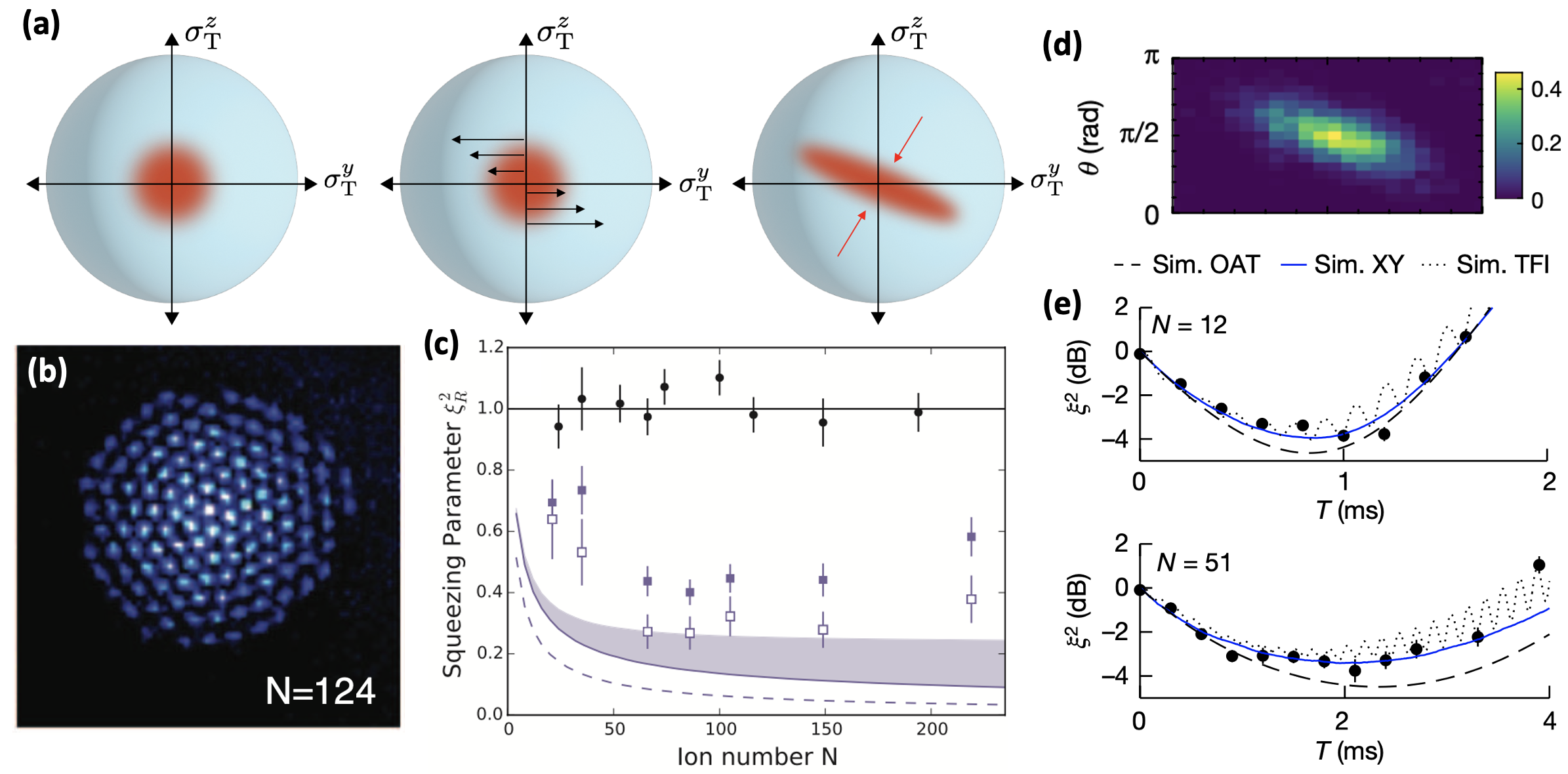}
\caption{{\bf Spin squeezing in 1D and 2D trapped ion quantum simulators.} 
{\bf a)} Schematic depicting the evolution of the quantum spin projection noise under a one-axis twisting Hamiltonian. (b) Image of a coulomb crystal of $N=124$ $^9$Be$^+$ ions in a Penning trap. (c) Ramsey squeezing parameter
measured for different 2D ensemble sizes $N$. The black points show data for the initial unentangled
spin state, while the solid purple squares show the lowest directly measured squeezing parameter with no corrections. (d) Measured Husimi Q-distribution of a 1D 12-ion spin chain for an interaction time where spin squeezing is observed. 
(e) Squeezing parameter as a function of time for a 1D chain of $N=12$ ions and $N=51$ ions. The time-scale where optimal squeezing is achieved is later for the larger spin chain, consistent with predictions, but the amount of spin squeezing does improve with system size, suggesting that the experiments are not in a scalable squeezing regime or that improvements are cutoff by decoherence. Panels (b,c) are adapted from \cite{Bohnet2016}, while panels (d,e) are adapted from \cite{franke2023quantum}. 
}
\label{fig:squeezing}
\end{figure}

In the previous section (Sec.~\ref{sec_high_energy}), the long-range nature of the ions' interactions was important for generating frustrated Ising couplings. 
Here, we discuss two other contexts where the power-law interactions of trapped ion systems are essential for realizing and exploring qualitatively distinct physics than can be realized in short-range interacting systems. 

\subsubsection{Quantum-enhanced metrology via spin squeezing}
\label{sec:spin_squeezing}

Quantum enhanced metrology makes use of entanglement to perform measurements with greater precision than would be possible using only classically correlated particles~\cite{giovannetti2011advances,pezze2018quantum}.
Perhaps the most well known example of a such a state is the so-called Greenberger–Horne–Zeilinger (GHZ) state, which features a macroscopic superpositions of all spins pointing up and all spins pointing down, $\psi_\textrm{GHZ} = (| \uparrow \cdots \uparrow \rangle + | \downarrow \cdots \downarrow \rangle)/\sqrt{2}$.
An $N$ spin GHZ state accumulates phase from an external magnetic field $N$ times faster than a single spin leading to a sensitivity that scales as $\sim 1/N$, the so-called \emph{Heisenberg} limit for sensitivity scaling; this contrasts with the more conventional \emph{standard quantum limit} scaling of $\sim 1/\sqrt{N}$, which is realized by an ensemble of uncorrelated sensors~\cite{giovannetti2011advances,monz201114,pogorelov2021compact}. 
The enhanced sensitivity of the GHZ state comes at a cost: if one loses even a single particle from the  state, the enhancement is lost, and thus the GHZ state is also significantly more sensitive to decoherence and particle loss. 
While this type of trade-off is an intrinsic feature of all metrologically-useful quantum states, certain forms of entanglement can be more robust to particular sources of noise and error~\cite{jeske2014quantum,zhou2018achieving}. 
This is precisely the case for \emph{spin-squeezed} states, where particle loss does not significantly degrade the sensitivity~\cite{kitagawa1993squeezed}. 
To understand the origin of the enhanced sensitivity associated with spin-squeezed states, it is helpful to recall one of the paradigmatic methods to generate such states (and which will be relevant for our discussion of trapped ion experiments below), namely, quench dynamics under the so-called one-axis twisting Hamiltonian~\cite{kitagawa1993squeezed,ma2011quantum},
\begin{equation}
\label{OAT}
H_\textrm{OAT}=\frac{1}{N} \sum_{i,j} J \sigma^z_i \sigma^z_j = \frac{(\sigma^z_T)^2}{N}.
\end{equation}
This model corresponds to all-to-all Ising interactions between $N$ spin-1/2 degrees of freedom and can naturally be recast in terms of a single ``large'' spin, $\sigma^z_T = \sum_i \sigma^z_i$.
Unlike the GHZ state, where the enhancement in sensitivity arises from faster phase accumulation, in the case of spin squeezing, the enhancement arises from a reduction in the spin projection noise of the underlying state~
\cite{toth2009spin,sinatra2022spin}. 
In particular, consider an initial product state with all spins aligned along the $x$-direction, $|\psi(t=0) \rangle = |\rightarrow \cdots \rightarrow \rangle$.
In order to detect a magnetic field in the $z$-direction, $H' = \sum B_z \sigma^z_i$, one can simply let $|\psi \rangle$ undergo Larmor precession for a set period of time and then measure the resulting phase accumulation. 
An intrinsic limit to the sensitivity of such an approach is given by the so-called spin projection noise, corresponding to the uncertainty of the quantum state in $yz$-plane [left, Fig.~\ref{fig:squeezing}(a)].
However, one can ``reshape'' this quantum uncertainty by evolving the product state under $H_\textrm{OAT}$~\cite{leroux2010implementation,braverman2019near}.
In particular, since $H_\textrm{OAT} \sim (\sigma^z_\textrm{T})^2$, the evolution causes the spin projection noise to be rotated by an amount proportional to $\sigma^z_\textrm{T}$; moreover, noise in the upper-half $yz$-plane becomes sheared in the opposite direction as the lower-half [middle, Fig.~\ref{fig:squeezing}(a)].
This leads to a reshaping of the spin projection noise (which nevertheless respects Heisenberg uncertainty) and can yield a direction where the noise is suppressed relative to the original product state [right, Fig.~\ref{fig:squeezing}(a)]. 
By rotating this axis to be perpendicular to the direction of the external field being sensed, one can achieve a sensitivity scaling as $\sim 1/N^{5/6}$, which is better than the standard quantum limit, but not quite at the Heisenberg limit. 
We note that there are other Hamiltonians, such as two-axis countertwisting, that do enable spin squeezing with Heisenberg-limited scaling~\cite{combes2004states}.

With these discussions in hand, it is clear that the tunable, long-range Ising interactions associated with trapped ions make them an ideal system for realizing and investigating spin squeezing dynamics. 

\vspace{2mm}

\noindent {\bf Spin squeezing in a 2D Penning trap}---In Reference~\cite{bohnet2016quantum}, the authors realize a 2D array of $^{9}$Be$^{+}$ ions [Fig.~\ref{fig:squeezing}(b)] in a Penning trap and encode a spin-1/2 degree of freedom using the $^2$S$_{1/2}$ ground state of the valence electron spin~\cite{crick2008two,britton2012engineered,mavadia2013control}.
Leveraging spin-dependent forces generated via a pair of detuned lasers (Sec.~\ref{sec_Hamiltonian}), Bohnet~\emph{et al.} implement a nearly all-to-all coupled Ising model with $H_\textrm{eff}=\sum_{i<j} \frac{J_0}{|i-j|^p} \sigma^{z}_i \sigma^{{z}}_j$ and $p$ varying from $0.02 \le p \le 0.18$.
The ultra-long-ranged nature of the power-law is reminiscent of spin squeezing experiments in atomic cavity QED setups~\cite{schleier2010squeezing,chen2014cavity,braverman2019near,leroux2010implementation}, where all-to-all interactions are generated via coupling to a cavity mode. 

In order to characterize the generation of spin squeezing, Bohnet~\emph{et al.} evolve an unentangled spin state, polarized along the $x$-axis (i.e.~$|\psi(0)\rangle$ from above) and measure the so-called squeezing parameter as a function of time
\begin{equation}
\label{squeezing_parameter}
\xi^2 \equiv N \frac{\min_{\hat{n}\bot \hat{x}}{\mathrm{Var}[\hat{n}\cdot \vec{\sigma_\textrm{T}}]}}{\langle \sigma^x_\textrm{T}\rangle^2 }.
\end{equation}
The numerator of $\xi^2$ captures the minimum variance in the $yz$-plane (i.e.~the axis with the smallest spin projection noise), while the denominator measures the remnant polarization in the $x$-direction. 
By independently measuring the decay of the spin polarization and the spin variance (as a function of an external rotation angle), Bohnet~\emph{et al.} demonstrate the generation of spin-squeezed states at short times. 
At longer time-scales, the spin variance begins to increase above its minimum value, as the  projection noise begins to wrap around the Bloch sphere. 
One of the central predictions for OAT-like dynamics is that the amount of spin squeezing improves as a function of system size $\xi^2 \sim 1/N^{2/3}$~\cite{kitagawa1993squeezed}.
To this end, the authors systematically vary the number of ions in their 2D Penning trap from $N=21$ to $N=219$, observing that the amount of achievable spin squeezing indeed improves until approximately $4.0 \pm 0.9$ decibels of spectroscopic enhancement  at $N=84$ [Fig.~\ref{fig:squeezing}(c)]. 
The plateau observed in $\xi^2$ for larger system sizes can be accounted for by a combination of photon shot noise, as well as elastic and spin-changing spontaneous emission.

\vspace{2mm}

\noindent {\bf Spin squeezing in a 1D Paul trap}---The ability to tune the power-law Ising interactions in trapped ions near the all-to-all coupled limit, allows for the direct realization of one-axis twisting dynamics (Eqn.~\ref{OAT}).
However, as one might imagine, it is challenging to scale up an interacting many-body system while maintaining uniform all-to-all interactions. 
This naturally leads to the question of whether shorter range interactions can still yield \emph{scalable} spin squeezing, with a sensitivity beyond the standard quantum limit.  
Recent theoretical advances suggest a positive answer to this question and have proposed a deep connection between spin squeezing and continuous symmetry breaking~\cite{comparin2022scalable,block2023universal,bornet2023scalable}.
In particular, if the effective temperature (i.e.~energy density) of an initial state is below the critical temperature for continuous symmetry breaking of certain model Hamiltonians, then the resulting quench dynamics are believed to generate spin squeezing~\cite{block2023universal,bornet2023scalable}. 
Perhaps the simplest example of such a setting is the ferromagnetic XY model, which can exhibit finite temperature easy-plane magnetic order with either nearest-neighbor interactions in 3D or power-law interactions in 1D and 2D. 

\begin{figure}[t]
\centering
\includegraphics[width=1.0\columnwidth]{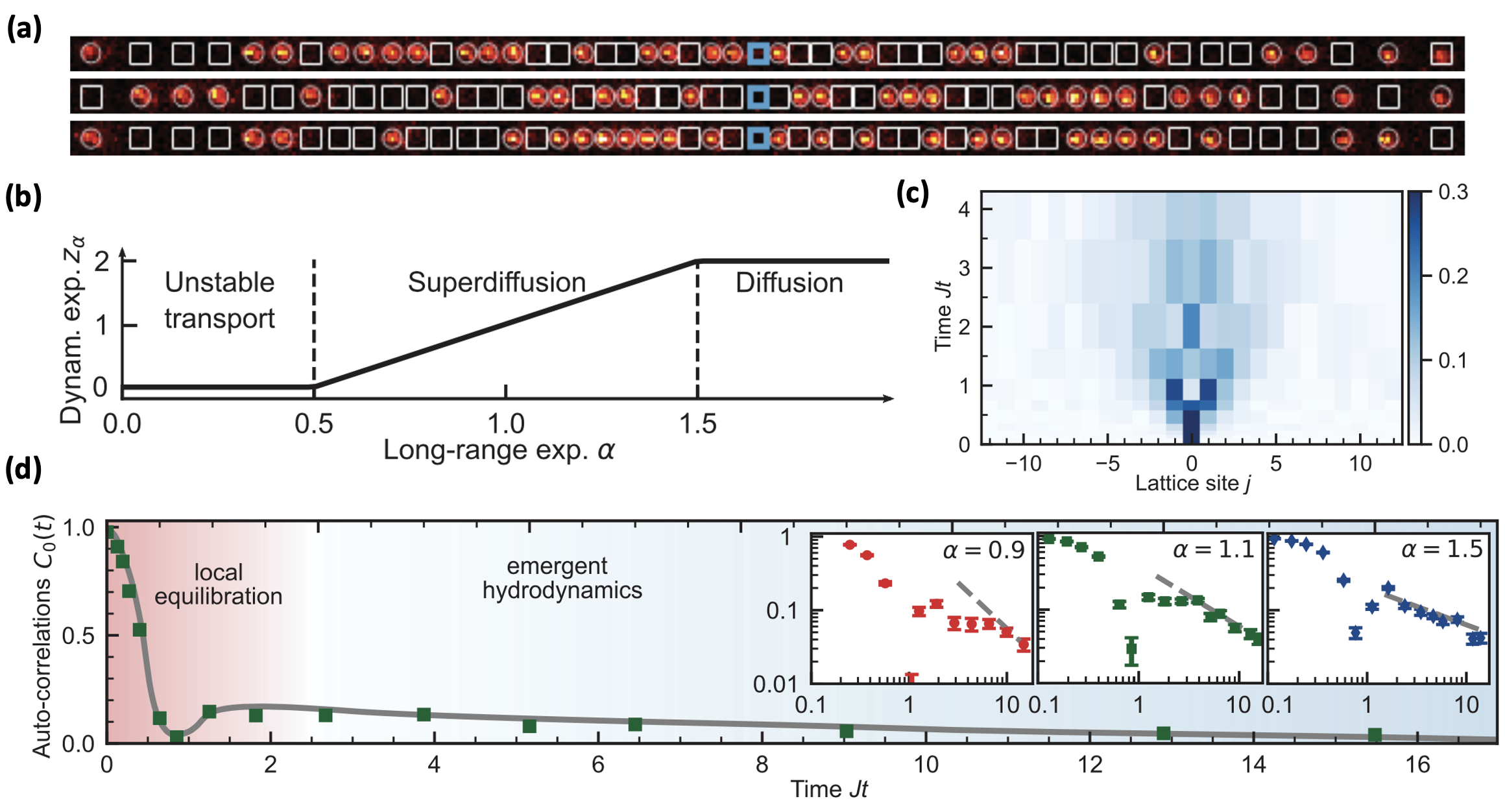}
\caption{{\bf Emergence of hydrodynamics in a 1D ion chain.} 
{\bf a)} In order to measured infinite temperature correlations, many different initial product states are averaged over. However, for each such state,  the central ion is deterministically prepared in the same state (blue box). Dark and bright spots indicate the two spin states of the ion chain. 
(b) Depicts different hydrodynamic universality classes characterized by the dynamical exponent $z$ as a function of the power-law exponent of the ion's spin-spin interactions.
(c) Spatiotemporal correlations of spin transport  at infinite temperature. 
The deterministically prepared excitation of the central ion slowly spreads through the system following the laws of classical hydrodynamics.
(d) Measured autocorrelation, $C_0(t)$
 for 51 ions and $\alpha = 1.1$. At short times
(red shading), the spin excitation quickly relaxes to a local equilibrium state.
At late times (blue shading), global conservation laws constrain the relaxation  leading to a slow power-law decay of the autocorrelations. (Insets) show the same autocorrelations on a double logarithmic scale for different values of
$\alpha$.  
Figure adapted from \cite{joshi2022observing}.
}
\label{fig:hydro}
\end{figure}

In Reference~\cite{franke2023quantum}, Franke \emph{et al.} trap a 1D chain (up to $N=51$) of $^{40}$Ca$^{+}$ ions in a  linear Paul trap and encode an effective spin degree of freedom in two electronic states, $|S_{1/2}, m=1/2\rangle$ and $|D_{5/2}, m = 5/2\rangle$.  
Again, as described in Sec.~\ref{sec_Hamiltonian}, spin-spin interactions are generated via a pair of laser that couples the spin states of the ions to their transverse
motional modes.
Much as above, the native Hamiltonian that emerges is a long-ranged Ising interaction. 
However, in the experiment of Franke \emph{et al.}, a strong driving field is applied which effectively creates a transverse magnetic field. 
For sufficiently strong driving, this transverse magnetic field dominates the Hamiltonian, and leads to an effective interaction (in the rotating frame of the drive), which precisely corresponds to a long-ranged XY model, 
\begin{equation}
H_{\textrm{XY}}=\sum_{i<j} \frac{J_0}{|i-j|^p} (\sigma^{+}_i \sigma^{{-}}_j + \sigma^{-}_i \sigma^{{+}}_j).
    \label{eq_XY_model}
\end{equation}
In principle, in a one dimensional system, it is predicted that relatively short-ranged power-laws with $\alpha \lesssim 1.5$ are sufficient to enable scalable spin squeezing. 
The experiments by Franke \emph{et al.} work with $p \approx 0.9$, which is longer-ranged than is strictly necessary for scalable squeezing and also leads to a super-extensive many-body spectrum. 
However, at the same time, the interactions are significantly shorter-ranged than either the OAT model or the nearly all-to-all interactions in the 2D setting described above. 
Starting from a polarized product state, Franke \emph{et al.} evolve the system under $H_\textrm{XY}$ and directly measure the  Husimi Q-distribution of a 12 ion chain.
At intermediate evolution times, they observe   shearing of the spin projection noise characteristic of spin squeezing [Fig.~\ref{fig:squeezing}(d)].
In addition, the squeezing parameter reaches its minimum value at earlier times for shorter ion chains [Fig.~\ref{fig:squeezing}(e)], consistent with theoretical expectations. 
However, the amount of spin squeezing that is achieved does not depend strongly on system size, yielding a maximum spectroscopic enhancement of approximately $3.2 \pm 0.5$ decibels. 
Somewhat analogous to the 2D case above, the
plateau observed in $\xi^2$ for larger ion chains can be accounted for by decoherence (i.e. finite $T_2$ time-scales); looking forward, the authors note that by doubling the coherence times of their system, they should be able to observe scalable spin squeezing up to the largest system sizes studied, i.e. $N=51$.

\subsubsection{Hydrodynamics with long-range interactions}

In an isolated many-particle quantum system, conventional wisdom holds that the late-time dynamics of conserved quantities usually exhibit an emergent classical description~\cite{esposito2005emergence,castro2016emergent,ye2020emergent}.
Understanding how to prove this fact and the precise way in which classical hydrodynamics emerges from  microscopic quantum interactions remain important open questions. 
On the experimental front, tremendous progress in time-resolved measurement techniques have enabled the direct observation of emergent classical \emph{diffusion} in several classes of short-range interacting quantum systems~\cite{sommer2011universal,moll2016evidence}.
The presence of long-range interactions can significantly alter the universality class of the emergent hydrodynamics, as power-laws enable the direct transport of spin excitations over many lattice sites~\cite{schuckert2020nonlocal,zu2021emergent}. 

Here, we focus on the example of the long-ranged XY model [Eqn.~\ref{eq_XY_model}] introduced in Sec.~\ref{sec:spin_squeezing}, which exhibits a $U(1)$ symmetry corresponding to the conservation of $\sigma^z_\textrm{T}$. 
The spin transport of this model is expected to depend sensitively on an interplay between the power-law $p$ and the dimensionality of the system. 
In particular, since $H_\textrm{XY}$ directly connects states with spin up at site $i$ and spin down at site $j$, $\left |\uparrow_i, \downarrow_j \right \rangle$, with the opposite configuration  $\left |\uparrow_j, \downarrow_i \right \rangle$, one can immediately use Fermi's Golden rule to compute a classical rate for spin exchange transitions, $W_{i \rightarrow j} = | \left \langle \uparrow_j, \downarrow_i \right  | H | \left |\uparrow_i, \downarrow_j \right \rangle|^2 \sim \frac{J_0^2}{|i-j|^{2p}}$.
To understand the role of this power-law, we note that (working in momentum space), in the presence of long-range interactions, the decay of a perturbation with wavevector $k$ can in general be written as $D k^2 + C_\textrm{LR} k^{\alpha - d} + Ck^4 +\cdots$, where $D$ corresponds to the spin diffusion coefficient, $\alpha$ is the effective classical transition rate, and $d$ is the dimension of the system~\cite{zu2021emergent}.
When $d < \alpha < d + 2$, the $k^{\alpha-d}$ term becomes the leading 
contribution to the dynamics and the system is no longer diffusive, entering instead, the so-called  Levy flight regime.

In Reference~\cite{joshi2022observing}, Joshi \emph{et al.} explore the emergence of hydrodynamics in the spin transport of a 1D ion chain, making full use of the platform's tunable power-law. 
As discussed above, the  long-range XY model leads to an effective classical spin exchange rate, $\alpha = 2p$ in the trapped ion system, suggesting that for $p > 3/2$, the spin transport reverts to diffusion, with a dynamical exponent $z=2$.
Meanwhile, for $1/2 \le p \le 3/2$, the system enters the Levy flight regime, where the spin transport is  described by random walks with long-distance jumps [Fig.~\ref{fig:hydro}(b)]. 
These long-distance jumps 
cause the spin transport to become anomalous and superdiffusive with  dynamical exponent, $z  =2p–1$~\cite{schuckert2020nonlocal}.

To probe the spin dynamics, Joshi \emph{et al.} create a spin excitation at time $t = 0$ in the center of the ion chain and track how it propagates in space and time (Fig.~\ref{fig:hydro}[c]).
The simplest such measure corresponds to the so-called survival probability, i.e.~the autocorrelation function of the central spin, $C_0(t) = \langle \sigma^z_0 (t) \sigma^z_0 (0) \rangle$.
Working with zero average total magnetization, Joshi \emph{et al.} investigate the infinite temperature survival probability by averaging the dynamics over multiple initial product states with the central ion deterministically prepared in the same state [Fig.~\ref{fig:hydro}(a)].
They observe that the dynamics of $C_0(t)$ exhibit  multiple distinct stages. 
First, the autocorrelation function exhibits rapidly damped oscillations corresponding to the approach to  local equilibrium.
 At later times, the system enters the hydrodynamic regime, where global equilibrium  is achieved through spin dynamics constrained by the conservation of total magnetization.
 In order to quantitatively characterize this regime for different power-law interactions, Joshi \emph{et al.} plot $C_0(t)$ on a double logarithmic axis [insets, Fig.~\ref{fig:hydro}[d]], where late-time hydrodynamics should in principle manifest as a power-law given by $-1/z$.
 Data are taken for power-laws in the Levy flight regime ($p=0.9, 1.1$) and at the boundary to the diffusive regime ($p=1.5$).
 While their time-scales are ultimately cutoff by finite system sizes, Joshi \emph{et al.} find that, at the latest achievable time-scales, their data are consistent with a $-1/z$ power-law exponent [insets, Fig.~\ref{fig:hydro}(d)].
 %
 In addition to studying $C_0(t)$, Joshi \emph{et al.} also attempt to extract spin transport coefficients by
 collapsing the full spatiotemporal profile of the spin correlations using hydrodynamic scaling functions predicted from Levy flights~\cite{joshi2022observing}.
 Finally, we note that trapped ion quantum simulations have also explored the opposite side of transport behavior where the dynamics of spin excitations are arrested owing to localization effects~\cite{smith2016many,morong2021observation}.




\section{Non-unitary dynamics from measurement}\label{sec_non_unitary}
Looking beyond purely-unitary dynamics, the ability to selectively measure subsets of qubits while retaining coherence in others allows one to synthesize arbitrary non-unitary quantum processes. Interspersing measurements with unitary Hamiltonian evolution or gates enables rapid preparation of certain long-range entangled states~\cite{tantivasadakarn2023hierarchy}, implementation of measurement-based models of quantum computation~\cite{briegel2009measurement}, active quantum error correction~\cite{terhal2015quantum}, and simulation of new types of measurement-induced phases and phase-transitions~\cite{potter2022entanglement,fisher2023random}, 

Trapped-ion architectures enable high-fidelity, low-cross-talk, mid-circuit measurements. In QCCD architecture, this can simply be done by shuttling the desired qubits to a measurement zone, well-separated from other qubits. Mid-circuit measurements are also possible in ion-chain architectures but require more complex shelving schemes to avoid measurement on selected qubits \cite{Erhard2021, Yang2022}.
In this section, we review progress in using mid-circuit measurements in trapped-ion QCCD architectures to implement topological phases of matter and perform active error correction, explore measurement-induced phase transitions, and implement compressed quantum simulation of large many-body models with small quantum memory.

\subsection{Preparing topological orders with measurement and feedback}
The dynamics induced by local unitary gates is constrained by fundamental speed-limits~\cite{lieb1972finite} that restrict the rate at which entanglement and correlations can be generated. 
The incorporation of measurements and adaptive operations based on long-range classical communication of measurement results (which is effectively instantaneous on present experimental length and time scales) can circumvent some of these speed-limits (see e.g.~\cite{tantivasadakarn2023hierarchy}). 

This section reviews the use of measurement-based circuits to prepare eigenstates of (topologically ordered) discrete lattice gauge theories. The  ground space and excitations of these can non-locally encode logical quantum memories in a way that is insensitive to local errors, and can therefore serve as the logical space of a quantum error correcting code. These states exhibit long-range entanglement and, using unitary gates, can only be prepared by deep circuits whose depth scales with the system size. For certain types of gauge theories, measurements and feedback (utilizing classical communication and unitary gates that adaptively depend on the measurement outcomes) can provide a shortcut to preparing ground-states in constant time (independent of the system size).
A classic example is that measurement of the stabilizers of the toric code~\cite{kitaev2010topological} (a lattice model of a $\Z_2$ gauge theory), followed by a single round of error correction, can prepare its ground state(s)~\cite{terhal2015quantum}.
In this section, we review recent experimental progress in trapped ion systems for implementing the ground-states of Abelian~\cite{iqbal2023topological,foss2023experimental} and non-Abelian~\cite{iqbal2024non} gauge theories with low-depth measurement-assisted circuits and demonstrating the anyonic braiding properties of excitations and defects of these systems.

\subsubsection{Lattice gauge theory and topological order}
In the experiments of Refs. \cite{iqbal2023topological,iqbal2024non}, it is useful to view the role of measurements as enforcing the Gauss' law of the gauge theory. 
For familiar electromagnetism, Gauss' law relates the divergence of electric field and charge density: $\nabla \cdot \vec{E} = \rho$. On a lattice, it is conventional to locate the charges, $\rho_i$, on sites, $i$, of the lattice and electric fields, $E_{ij}=-E_{ji}$ (canonically conjugate to the vector potential $A_{ij}$), on (directed) links of the lattice connecting neighboring sites $i,j$. Quantization of charge dictates that $\rho_i$, and hence also $E_{ij}$, are integer-valued. 

For a discrete, $\Z_2$ lattice gauge theory, the charges and electric fields can only take binary values $\rho_i,E_{ij}\in \{0,1\}$. To encode these into qubits living on the sites (with Pauli matrices $\{\sigma^{x,y,z}\}$) and links (with Pauli matrices $\{X,Y,Z\}$), we can identify $(-1)^{\rho_i} = \sigma^z_i$ and $(-1)^{E_{ij}} = X_{ij}$, and $(-1)^{A_{ij}} = Z_{ij}$. In this discrete gauge theory, exponentiation of the Gauss law equation gives: 
\begin{align}
    \prod_{\<ij\>\in+_i} X_{ij} = \sigma^z_i
    \label{eq:z2gauss}
\end{align}
where $+_i$ denotes the set of links emanating from site $i$.

\begin{figure*}[t]
\includegraphics[width=1.0\textwidth]{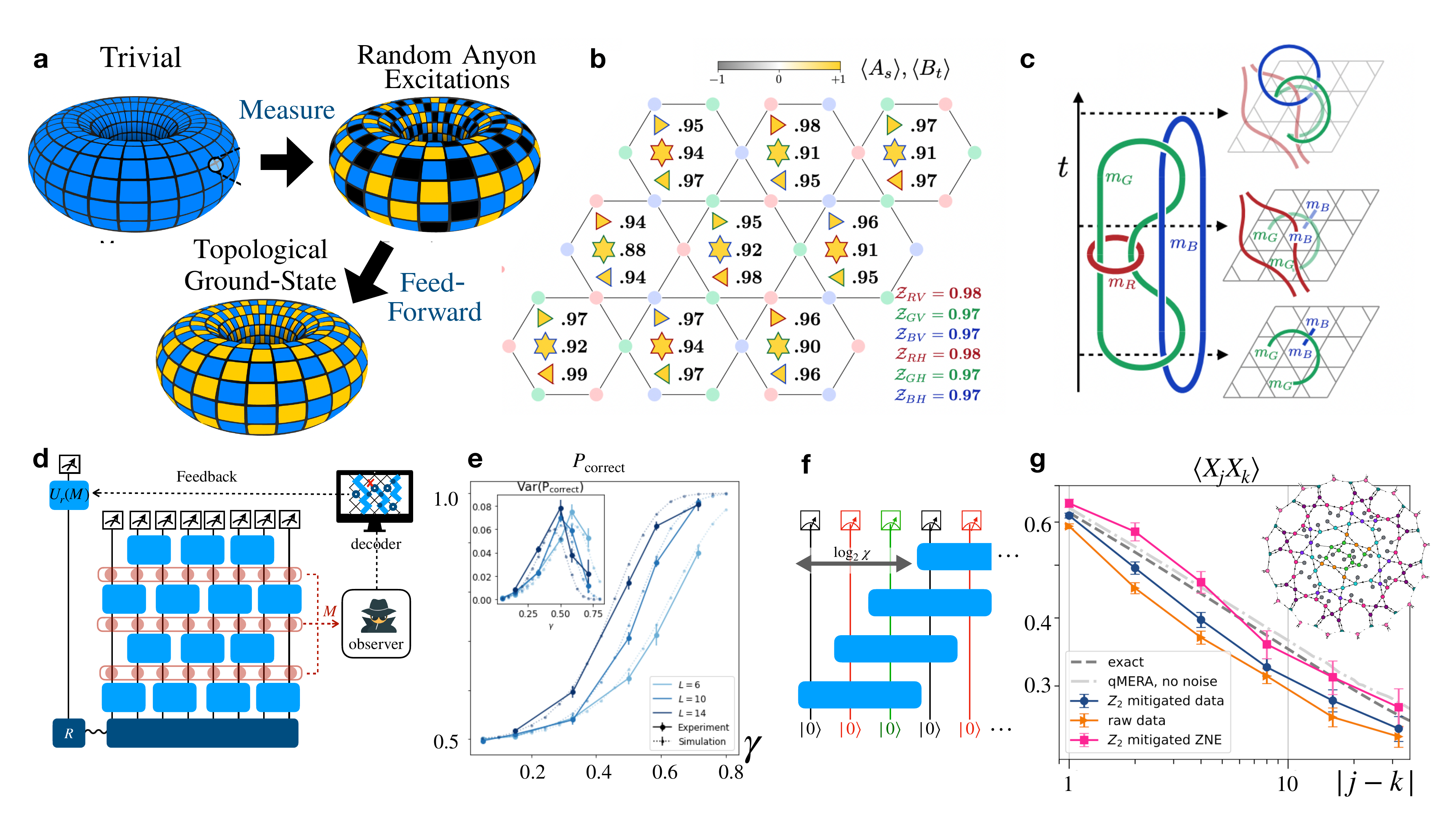}
\caption{
{\bf Creating and simulating highly entangled states and criticality with measurements and feed-forward.}
{\bf a)} Schematic of preparation of topological order via measurements and feed-forward. 
{\bf b)} The experiment in Reference \cite{iqbal2024non} adapted this procedure to create non-Abelian topological order. Measurements of ground-state projectors confirm high overlap with the ideal topological order state. {\bf c)} Braiding of three $m$ particles in the fashion of Borromean rings gives a non-trivial phase that can be detected via an interferometric protocol, revealing the non-Abelian statistics of these quasiparticles. {\bf d)} Schematic of ``decoding" setup for observing measurement-induced phase transitions (MIPTs) and {\bf e)} finite size scaling evidence for an observable-sharpening phase transition in $L=6,10,14$ length ion chains. $P_{correct}$ denotes the probability that the decoding algorithm assigns to the correct charge. The presence of a phase transition at a critical measurement strength $\gamma=\gamma_c\approx 0.4$ is signaled by steepening crossover in the probability, $P_{correct}$, that the decoding algorithm predicts the correct charge and a sharpening peak in the variance of this quantity (inset).
\textbf{f)} Generation of a matrix product state (MPS) via a sequential quantum circuit (SQC).  Qubit reuse enables the state output by an SQC to be sampled using a number of qubits proportional to the width of the staircase. \textbf{g)}.  Tree-like tensor networks, such as the multi-scale entanglement renormalization ansatz (MERA) also admit significant compression via qubit reuse (inset). In this way, Reference\,\cite{haghshenas2023probing} used 20 physical qubits to represent a 128 site critical spin chain and accurately extracted scaling exponents. Panel (a) adapted from Ref.~\cite{iqbal2023topological}. Panels (b,c) adapted from Ref.~\cite{iqbal2024non}. Panels (d,e) adapted from Ref.~\cite{agrawal2023observing}. Panel (g) adapted from Ref.~\cite{haghshenas2023probing}.}
\label{fig:measurement}
\end{figure*}

\subsubsection{Preparing Toric Code ($\Z_2$ Gauge Theory)}
The standard toric code state preparation procedure proceeds as follows: First, one starts all the link qubits in a $Z_{ij}=+1$ product state. In the gauge theory language, this corresponds to initializing with zero magnetic flux through all plaquettes of the lattice. Then, an ancilla qubit is entangled with the value of $\prod_+ X$ on links that end at site $i$, such that measuring $\sigma^z$ of the ancilla measures the $\prod_+ X$ stabilizer of the toric code, without measuring the individual qubits in the stabilizer. Reference~\cite{tantivasadakarn2021long} showed that this procedure can be usefully re-interpreted as ``gauging" (i.e., enforcing Gauss' law) by measurement. Namely, if we view the ancilla qubit as the charge qubit and repeat this for every site in the lattice, this procedure prepares a state that satisfies Gauss' law and results in a random (dependent on measurement outcome) arrangement of $\Z_2$-electric gauge charges. This is a random excited state of the gauge theory, but it can be mapped into a ground-state by applying strings of $Z_{ij}$ operators to pairwise annihilate all the measured electric gauge charges. 

The result of this procedure is to prepare a ground-state (vacuum) of the $\Z_2$ gauge theory (which reduces to the standard toric code~\cite{kitaev2010topological} if one discards the ancilla qubits) with no electric charges or magnetic fluxes. 
When the lattice is a torus with periodic boundary conditions, there are actually four distinct ground-states that can be labeled by the $\Z_2$-gauge magnetic flux through each cycle of the torus. This ground-space can serve as a logical space of an error correcting code, since no local operator can cause transitions (errors) between the different ground states. 

Of course, in planar fabricated qubits, creating a torus is pure fantasy. However, in architectures that allow qubit shuttling, it becomes an experimental possibility.
 Reference~\cite{iqbal2023topological} implemented this adaptive measurement procedure to prepare toric code ground states on a $4\times 4$ square grid was using 16 ion qubits in Quantinuum's QCCD quantum processor. The state preparation was validated by measuring errors in the plaquette stabilizers $\<X,Z^{\otimes 4}\>$ in the $3-10\%$ range, and errors in expectation values of logical $X$ and $Z$ loop operators in the $3-9\%$ range. 
Direct experimental evidence of long-range entanglement was obtained by measuring a 2-Renyi version of the topological entanglement entropy (TEE)~\cite{levin2006detecting, kitaev2006topological}, $\gamma$ defined by the entanglement scaling form: $S_A \approx \alpha|\d A|-\gamma$, where $|\d A|$ is the boundary of region $A$, $\alpha$ is a non-universal constant, and the ideal value for $\gamma$ is $\log 2$. The universal constant $\gamma$ can be extracted by considering an appropriate linear combination of entanglements for overlapping regions~\cite{levin2006detecting, kitaev2006topological}. In the trapped-ion experiment, this was done using $2\times 2$ to $2\times 3$ qubit regions, resulting in $\gamma/\log 2\approx 1$ with typical errors of $5-7\%$.

The experimental demonstration of~\cite{iqbal2023topological} was also able to demonstrate another striking feature of this toric code model: that lattice dislocations act as non-Abelian twist defects. Specifically, dragging an $e$ particle around such a defect converts it into an $m$ particle.
These twist defects have non-Abelian properties identical to Majorana fermion bound-states long sought in topological superconductors; in particular, each pair of such defects non-locally encodes a logical qubit~\cite{bombin2010topological} in a manner that is protected from local decoherence.

\subsubsection{Preparing a non-Abelian gauge theory}
Whereas Abelian gauge theories can store information only in the ground states and defects, non-Abelian topological orders can also encode logical quantum information in particle-like excitations~\cite{nayak2008non}. $N$ non-Abelian particles of type $a$, non-locally encode a logical code space of size $\sim d_a^N$, where $d_a$ is a property of the anyon known as its quantum dimension. The non-local encoding cannot be corrupted by local noise, and hence the code space forms a good quantum memory. Further, the information stored in this memory can be manipulated by braiding the non-Abelian anyons around each other, which performs unitary operations (gates) on the code space that depend only on the topology of the rearrangement of the anyon world lines and are also fault-tolerant.
In sufficiently rich non-Abelian topological orders (codes), this braiding can implement universal, fault-tolerant topological quantum computation~\cite{nayak2008non}.

Reference~\cite{tantivasadakarn2023shortest} showed that short-depth measurement circuits and adaptive feedback could be used to prepare the ground-space of certain classes of discrete gauge theories with non-Abelian excitations. 
A recent experimental realization~\cite{iqbal2024non}, used this protocol to prepare a non-Abelian topological order corresponding to a
discrete non-Abelian gauge theory with gauge group $D_4$ (the dihedral group of order $8$, i.e., the symmetry group of a square), and to confirm the non-Abelian statistics of its excitations.
The anyons of the $D_4$ gauge theory can be labeled by three ``colors" [red (R), green (G), and blue (B)] of gauge charge and flux. The $\Z_2$-valued electric gauge charges: $e_{R,G,B}$ are Abelian particles with bosonic self-statistics. The non-Abelian gauge fluxes $m_{R,G,B}$ are non-Abelian particles with quantum dimension $d_m=2$, i.e. each can store a (non-locally encrypted) logical qubits' worth of information. 

A useful perspective on the $D_4$ gauge theory~\cite{propitius1995topological} that enabled its experimental implementation is that $D_4$ gauge theory can be viewed as three copies of a simpler $\Z_2$ gauge theory (one for each R,G,B color), but twisted together by a topological term that does not affect the electric charge sector, but gives the magnetic fluxes their non-Abelian properties. 
Reference~\cite{iqbal2024non} used this perspective to implement a 36-qubit lattice realization of the $D_4$ gauge theory on a Kagome net formed from three inter-penetrating triangular lattices (one for each $R,G,B$ color), with 27 qubits on the links representing the gauge magnetic fields, and 9 plaquette qubits to represent gauge electrical charges (physically implemented by 3 ions using qubit reset and reuse techniques).
The gauge-field qubits are first ``twisted" by a short-depth unitary circuit, $U_{\rm SPT}$, to convert the gauge group from three copies of $\Z_2$ to $D_4$. Then, the Gauss' law from the $D_4$ gauge theory was enforced on every site by measurement, followed by a single round of error correction to remove stray electrical charges. 
By ensuring that no non-Abelian particles are present in the measurement round, the error-correction step can be done deterministically in a single shot (general conditions for single-shot preparation are discussed in~\cite{tantivasadakarn2023shortest}).

The fidelity of the state preparation was validated by measuring ground-state projectors, which lower-bounds the total many-body state fidelity $\gtrsim 75\%$, corresponding to $\geq 98\%$ fidelity per qubit.
The non-Abelian nature of the $m_{R,G,B}$ anyons in this state was then probed by interferometrically measuring the topological phase obtained by braiding the world-lines of $m$ particles of each three types into a knot called a ``Borromean braid" (see Fig.~\ref{fig:measurement}), which gives a non-trivial phase if and only if the anyons have non-Abelian exchange statistics.
The Borremean braid phase was measured interferometrically using a Hadamard test circuit to be within $2\%$ of the theoretical value of $\pi$, directly experimental confirming their non-Abelian character.

\subsubsection{Outlook}
These experiments offer promising proof-of-principle demonstrations of key capabilities required for topological quantum error correction. Two main challenges remain: First, these experiments do not yet perform the repeated rounds of error-correction needed to stabilize the topological phase as a fault-tolerant quantum memory. Second, the topological operations in both examples do not permit universal quantum computation, requiring additional ingredients such as code switching, magic-state distillation.
A potentially promising work-around would be  to prepare a more complicated $S_3$ (the permutation group with three elements) gauge theory, which can be done with only two rounds of measurement and feedback~\cite{bravyi2022adaptive,tantivasadakarn2023hierarchy}, and would enable universal topological computation via a combination of measurement and anyon braiding~\cite{mochon2004anyon}.

Further, while topological codes encode physical qubits into logical qubits at an asymptotically optimal rate for 2D qubit arrays, non-local virtual connectivity afforded by qubit shuttling principle enables access to codes that cannot be embedded in 2D (or even $3d$), which may have dramatically enhanced code properties, requiring (asymptotically) far fewer qubits to encode a desired number of logical qubits and achieve a desired rate of error suppression~\cite{gottesman2013fault}.

\subsection{Measurement-induced phase transitions}
The ability to selectively measure qubits during a quantum circuit enables access to a new regime of monitored quantum dynamics, in which the system undergoes non-unitary dynamics (due to measurements) but remains in a pure quantum state ``trajectory", $|\psi_M\>$ that depends on the (stochastic) measurement outcomes, $M$. 
These capabilities enable one to tune the effective strength of the coupling, $\gamma$, of the system to the measurement apparatus and probe the detailed evolution of the measurement-induced ``collapse" of quantum superpositions
For a microscopic system, such as a single qubit, there is no sharp distinction between the weak measurement regime, where quantum superpositions are only weakly perturbed, and strong measurement regimes, where they fully collapse. These regimes are connected by a smooth crossover, with analytic dependence of all measurable quantities on measurement strength. By contrast, theoretical studies (see \cite{potter2022entanglement,fisher2023random} for reviews) show that in the limit of large system sizes ($L\rightarrow\infty$) and long measurement times ($t\rightarrow \infty$), this crossover can sharpen into a sharp \emph{measurement-induced phase transition (MIPT)} separating distinct weak- and strong- measurement phases.

Models of MIPTs typically consider monitored quantum circuits with randomly chosen gates $U$ drawn from a distribution $P(U)$ to ensure the unitary evolution rapidly and chaotically generates entanglement. Variable measurement strength, $0\leq \gamma\leq 1$, with $\gamma=0$ corresponding to no measurement, and $\gamma=1$ corresponding to a strong projective measurement, can be accomplished either by 1) weakly measuring each of the system qubits with strength $\gamma$ (utilizing an ancilla qubit that is that becomes partially entangled with a system qubit before being projectively measured), or 2) by projectively measuring a randomly chosen fraction, $\gamma$, of qubits. 

The originally-proposed MIPT~\cite{li2019measurement} occurs as a phase transition in a 1D qubit array of length $L$ between i) a scrambling phase ($\gamma>\gamma_c$) whose trajectories, $|\psi_M\>$ rapidly develop entanglement at a finite rate that saturates to a volume law at late times, e.g., the half-system bipartite entanglement of a system of size $L$ scales asymptotically as $S(t,L)\sim {\rm min}(t,L)$, and ii) a disentangling phase ($\gamma<\gamma_c$) where single-qubit measurements suppress the entanglement resulting in $S(t,L)\sim {\rm constant}$. A host of other MIPTs have then been theoretically proposed in related models, including those where measurement induces distinct types of quantum phases with different symmetry-breaking or topological order in trajectories~\cite{fisher2023random}.

A fundamental obstacle to observing this change in trajectory entanglement is that the transition is only observable in individual trajectory $|\psi_M\>$, and \emph{not} in the ensemble-averaged state $\sum_M |\psi_M\>\<\psi_M|$. Multiple copies of a state $|\psi_{M}\>$ are required to estimate entanglement (or any other observable capable of revealing the transition). Since the measurement records, $M$, are random (according to Born's rule), it is exponentially unlikely (in $t \times L$) to repeat. This leads to the so-called post-selection problem: incurring a prohibitively large, $\sim e^{Lt}$, sampling overhead.
Various alternative means to observe measurement-induced phase transitions without post-selection have been proposed~\cite{potter2022entanglement,fisher2023random}. Each involves interaction between the monitored circuit and an observer who performs a ``decoding" computation to process the measurement outcomes. The output of this decoder, $D(M,U)$ is then fed back into the system such that the results can be checked in a single shot for each trajectory, evading the post-selection problem.
In these decoding and feedback settings, the existence and critical measurement strength of an MIPT depends jointly on the dynamics of the monitored quantum system and the decoder algorithm, closely analogous to quantum error correction threshold transitions.
Experimental realizations are further simplified by initializing the system in an entangled state with a single reference ancilla qubit, $R$. Depending on the phase of the monitored dynamics, the evolution either disentangles $R$ or preserves the coherent information between $R$ and the system. 
In this setup, instead of probing the many qubit properties of the system, one needs only to ``decode" the reference, $R$, to observe the transition.
Below, we review two trapped-ion experiments that leverage this alternative ``decoding" perspective to observe finite-size evidence for the presence of MIPTs.

\subsubsection{Purification transition}
Reference~\cite{noel2022measurement} implemented a small-scale demonstration of an MIPT in a chain of trapped $\rm ^{171}Yb^+$ ions undergoing a monitored circuit with randomly chosen Clifford gates interspersed by measurements. In the strong-measurement phase, the reference ancilla, $R$, will be purified (disentangled from the system). Thus, there exists a single-qubit recovery operation $U_r(M)$ that can rotate $R$ to a fixed state, $|0\>$. The success probability of this recovery operation (determined simply by measuring whether $R$ is in state $|0\>$) then serves as an order parameter for the entanglement MIPT.  
Exploiting the classical simulability of Clifford circuits, Reference~\cite{noel2022measurement} designed a decoder to identify $U_r$ based on the measurement outcomes, $M$, and observed finite-size behavior consistent with the expectation of two distinct (weak and strong) measurement phases. 
In this experiment, mid-circuit measurements (which are challenging in ion-chain architectures) are avoided by writing the measurement outcome onto ancillary ``measurement" qubits that are measured only at the end of the circuit. Further, the decoding and recovery operation was implemented entirely through quantum logic within the quantum circuit. The use of deferred measurements strongly limited the range of accessible circuit volumes for this experiment. Nevertheless, it was possible to design a model with small enough critical measurement rates to view a qualitative change in the recovery success probability as a function of the number of measurements in circuits with up to $L=6$ qubits and up to $|M|=3$ measurements.

\subsubsection{Observable sharpening transition}
Observable-sharpening transitions are a related but distinct class of MIPT separating weak (fuzzy) and strong (sharp) measurement regimes characterized by whether an initial superposition state with quantum uncertainty in the value of an observable, $O$, collapses under measurements (becomes ``sharp"), or remains fuzzy and uncertain. This MIPT class can be equivalently framed as whether a reference ancilla, $R$, entangled with different $O$ eigenstates, gets disentangled (sharp phase), or remains entangled (fuzzy phase) by the monitored dynamics. 
Here, in order to retain an invariant meaning through the circuit evolution, it is natural to restrict to quantum non-demolition measurements, the simplest class of which deal with observables $O$ that are conserved by the monitored dynamics.
The fluctuations of $O$ in a given trajectory $|\psi_M\>$ serve as an order parameter for this MIPT, but like entanglement, these require post-selection to observe. 
An alternative perspective of this transition that evades post-selection issue~\cite{potter2022entanglement} asks whether an observer learns (or not) enough information from the measurements to successfully predict the value of $O$. The prediction can then be checked in a single-shot by comparing the prediction against the measured value of $O$. In the sharp phase, an optimal decoding algorithm can successfully predict $O$ 100\% of the time, whereas in the fuzzy phase, no decoder can do better than random guessing.

Reference~\cite{agrawal2023observing} implemented an experimental demonstration of an observable sharpening MIPT in Quantinuum's system module H1-1 trapped ion processor. The experiment explored the monitored dynamics 1D array of qubits. The observable in question was the total ``charge" $Q=\sum_i  (2Z_i-1)$ of the qubits, and the monitored circuits consisted of Haar-random charge-conserving two-qubit gates and local weak measurements of $\Z_i$ of tunable strength $\gamma$ (performed using an ancilla qubit). 
Multiple decoders were used to predict the charge, including: an optimal decoder based on (non-scalable) brute-force classical simulation of the monitored circuits and a ``stat-mech" decoder that represents the optimal decoder for the scenario where the observer is ignorant of the precise gates, $U$, used in the circuit. Crucially, the stat-mech decoder observes a charge-sharpening transition at a slightly higher measurement rate than the optimal decoder but could be efficiently computed by matrix-product state (MPS) techniques even for generic classes of circuits that cannot be classically simulated (e.g., non-Clifford circuits, in the volume-law entangled regime of the monitored dynamics).
Finite-size evidence for the observable sharpening transition was observed in the success and variance of the decoders predictions, for a sequence of system sizes from  $L=6,10,14$ for circuit depths $t=L/2$~(Fig.~\ref{fig:measurement}).
To suppress noise, mid-circuit measurements were processed via the dual stat-mech model to locally detect errors that could not be detected by operations performed at the end of the circuit.

\subsubsection{Outlook}
The theoretical developments in studying MIPTs and related concepts suggest that statistical mechanics (stat-mech) concepts such as phase transitions, universality, and critical exponents can arise in quantum circuits, and stat-mech tools have already revealed examples of sharp phase transitions in quantum error correction, quantum communication channel capacity, and the complexity of classical simulations of quantum systems~\cite{potter2022entanglement,fisher2023random}.
These insights have intrinsic theoretical value for understanding the capabilities and limitations of quantum computation, but the above examples show that certain classes of them may also be observable directly in experiments.
It remains an open question which classes of these transitions can be effectively observed in experiment, i.e., without post-selection and using efficient classical (or quantum) computations to perform any necessary decoding and feedback.

\subsection{Quantum tensor network methods}
While a general state of $N$ qubits can have $O(N)$ entanglement entropy, quantum states that arise in many physically relevant contexts (e.g., low energy states of local Hamiltonians) have entanglement entropy $S\ll N$. Classical tensor network methods attempt to efficiently parameterize such quantum states, ideally in ways that are amenable to efficient classical manipulations.  If such efficiencies are achievable via suitable classical representations, it is natural to ask: Does a \emph{quantum} representation of the state of $N$ qubits truly require $N$ qubits, or can we find a way to represent the state with $n\sim S\ll N$ qubits? Quantum tensor network methods are an attempt to do exactly that.

To illustrate some of the key ideas behind quantum tensor network methods, consider the simplest and best-known TN ansatz: the matrix-product state (MPS).  As originally pointed out (using the slightly different language of cavity-QED) in \cite{schon2007sequential}, any matrix product state can be prepared as a sequential quantum circuit (SQC) with a staircase structure, as shown in Fig.\,\ref{fig:measurement}(a).  By letting the blue unitaries act on $\log_2\chi$ adjacent qubits, \emph{any} bond-dimension $\chi$ MPS can be represented at the output of the circuit.  The causal structure of an SQC enables the outputs to be measured with a remarkable economy of qubits.  Consider that prior to the application of the second (from bottom) blue gate, the leftmost (black) qubit has already reached the end of the circuit and can be measured.  By resetting that qubit to the $\ket{0}$ state and reusing it as the fourth qubit, the second blue gate can be executed, after which the second (red) qubit is measured and liberated for reuse.  Iterating this procedure to the right, the entire circuit can be executed using only 3 qubits (black, red, and green), which are reused to represent a state of arbitrary spatial extend via the repeating color pattern shown in the figure.  Note that the number of qubits required in this procedure ($n=3$ in the example here) is set by the width of the unitaries: $n=\log_{2}\chi$. Because states with entanglement entropy $S$ can generally be represented accurately as MPS with bond dimension $\chi\sim 2^{S}$, we find that only $n\sim S$ qubits are required to represent such states, as speculated above.  
Crucially, compression of an SQC via qubit reuse requires carrying out both measurement and state initialization in the middle of a circuit with minimal cross-talk on the un-measured qubits.

In general, implementations of MPS via qubit reuse can be viewed as physical implementations of the quantum channel that any MPS induces on the internal/bond degrees of freedom when they are viewed as indexing states in a (fictitious) Hilbert space. Here, the constant resetting and reuse of qubits is responsible for the non-unitary nature of that channel (and the existence of a steady state in the case of a translationally invariant MPS). From this perspective, the spatial extent of the MPS is mapped onto the temporal extent of that quantum channel, effectively hiding one spatial dimension of the underlying physical system being represented in the time direction of a quantum circuit.  This point of view inspired the original language of ``holographic simulation'' \cite{kim2017holographic} to describe the combination of SQCs (for representing limited entanglement states) with qubit reuse.  The connection to quantum channels also enables direct probes of the entanglement properties of an infinite state to be made by studying the density matrix of the small number of physical qubits representing it via qubit reuse \cite{gopalakrishnan2019unitary}. This connection was used in Reference~\cite{foss2022entanglement} to experimentally measure clear signatures of the expected divergence of entanglement entropy at a quantum phase transition (in an infinite system) using only three physical qubits.

%
%
%

%

The connection between SQCs and classical tensor networks goes considerably deeper than the MPS example given above. This type of compressed quantum representation of limited entanglement states can also be applied to 2D, or even 3D, systems \cite{kim2017holographic,foss2021holographic}, and in the setting where $\chi\sim 2^S\sim 2^n$ is too large to admit classical calculations it in principle enables variational optimization of tensor network states using a quantum computer \cite{kim2017robust,liu2019variational,foss2021holographic}.  In fact, a large class of 2D tensor networks introduced in Reference\cite{PhysRevLett.128.010607}, termed plaquette-PEPS, can be formulated as SQCs and admit significant compression via qubit reuse.  Such states encompass isometric TNS~\cite{PhysRevLett.124.037201}, which capture a variety of non-trivial 2D states including those representing long-range-entangled topological orders~\cite{soejima2020isometric}.

Moreover, tree-like tensor networks such as the multi-scale entanglement renormalization group (MERA) \cite{PhysRevLett.101.110501}, which can be used to prepare highly-entangled quantum critical ground-states, can also be recast as SQCs with thickness proportional to the tree-depth $(\sim \log L)$. This representation was exploited in Refs.~\cite{anand2023holographic,haghshenas2023probing} to perform quantitative calculations of universal power-law scaling of correlations at the 1D Ising critical point (and non-integrable deformations thereof~\cite{anand2023holographic}) on long chains of up to 128 sites using only 20 trapped ion qubits.

Each example of a quantum tensor network method given above begins with an approximate/compressed classical tensor network ansatz and extends it into a quantum circuit in which the original classical efficiencies of the ansatz are expressed as qubit number efficiencies. However, the idea of compressing a circuit via qubit reuse can be considerably generalized and unified \cite{decross2023qubit}, and is closely related to the rapidly developing field of general purpose optimization methods for the contraction of tensor-networks (see for example Reference\,\cite{gray2021hyper} and references therein).  The key concept underlying qubit reuse methods is the \emph{past causal cone} associated with a given output qubit, which is the collection of all gates and input qubits in the circuit that can causally impact a measurement on the output qubit.  Given sufficient qubits to initialize all such inputs, the gates can be executed and the output qubit can be measured, reset, and reserved for later use. Another output is then identified, and all input qubits in its causal cone (excluding any already initialized in the previous causal cone) are initialized using one less qubit than naively expected by reusing the already measured qubit. How many qubits are ultimately required to execute an entire circuit in this fashion depends on how effectively one can identify successive output qubits whose past causal cones require a small number of additional qubits to fully initialize.

Circuit compression via qubit reuse can be understood as the problem of identifying the optimal measurement order for the output qubits of a quantum circuit in order to minimize the total number of qubits, $M$ required to sample the full output of an $N>M$ qubit circuit via the qubit reuse method described above. This method, in turn, can be viewed as an attempt to contract the tensor network representing the circuit with the smallest possible memory footprint $\sim 2^M$, \textit{subject to the constraint of causal ordering} of tensor contractions.

These general considerations unify a variety of tensor-network-based qubit reuse methods that have been used to efficiently represent tensor network states time-evolved by brickwork circuits. In particular, any circuit with geometrically local gates on $L^d$ qubits and finite depth $\sim T$ can be redrawn as a SQC with thickness $\sim T\times L^{d-1}$, which can be utilized~\cite{napp2022efficient,foss2021holographic}  to simulate Trotterized non-equilibrium dynamics starting from a correlated initial state. For example, Reference~\cite{chertkov2022holographic} used these techniques to quantitatively model the quantum chaotic dynamics of a kicked-Ising model in a chain of $32$ qubits using only $\sim 8-10$ trapped ions in Quantinuum's system model H1 processor and observed characteristic features of quantum chaos, and correlations confined to light cone boundaries at special dual-unitary points in the model. Similar qubit reuse techniques were also used to measure critical properties in a non-equilibrium absorbing-state phase transitions~\cite{chertkov2023characterizing}, realizing a depth 72 circuit on 73 qubits using only 20 physical trapped-ion qubits.  While the previous two examples had simple enough causal structure for optimal qubit reuse based compression to be identified empirically, optimal qubit reuse can be formulated as the optimization version of a constraint programming and satisfiability (CP-SAT) problem, and solved exactly in an automated way for circuits of moderate size. This approach, along with heuristic adaptations that scale well to arbitrarily large circuits \cite{decross2023qubit}, has been used to compress shallow circuits based on random graphs in order to implement a 130 qubit QAOA algorithm on only 32 qubits \cite{moses2023}.

\section*{Outlook}
The selected topics reviewed above highlight only a fraction of the interesting recent developments in trapped-ion quantum simulators and processors. Our review focuses on applying hardware capabilities such as flexible connectivity or long-range interactions, and employing mid-circuit measurements and qubit reuse to enhance quantum simulation capabilities, and to experimentally access new regimes of quantum dynamics.
The scale and complexity of these systems are beginning to challenge classical simulation. An important outstanding task will be to understand how to continue improving qubit and gate operations, scaling the size and complexity of trapped-ion architectures, advancing quantum simulation algorithms to solve practical science and technology problems that lie beyond the reach of classical supercomputers, and to devise methods to suppress or actively correct errors to build towards fault-tolerant operations.

\section*{ACKNOWLEDGMENTS}
The authors gratefully acknowledge Marcello Dalmonte, Ruben Verresen and Zohreh Davoudi for a careful reading of the manuscript and many helpful suggestions. This material is based upon work supported by the U.S Department of Energy, Office of Science, Office of Nuclear Physics under the Early Career Award No. DE-SC0023806, the NSF CAREER Award (grant No. PHY-2144910), the Office of Naval Research Young Investigator Program (grant no. N00014-22-1-2282), and the NSF STAQ II program. 

\bibliographystyle{naturemag.bst}
\bibliography{qs_review}

\begin{thebibliography}{100}
\expandafter\ifx\csname url\endcsname\relax
  \def\url#1{\texttt{#1}}\fi
\expandafter\ifx\csname urlprefix\endcsname\relax\def\urlprefix{URL }\fi
\providecommand{\bibinfo}[2]{#2}
\providecommand{\eprint}[2][]{\url{#2}}

\bibitem{Bruzewicz2019}
\bibinfo{author}{Bruzewicz, C.~D.}, \bibinfo{author}{Chiaverini, J.},
  \bibinfo{author}{McConnell, R.} \& \bibinfo{author}{Sage, J.~M.}
\newblock \bibinfo{title}{{Trapped-ion quantum computing: Progress and
  challenges}}.
\newblock \emph{\bibinfo{journal}{Applied Physics Reviews}}
  \textbf{\bibinfo{volume}{6}}, \bibinfo{pages}{021314} (\bibinfo{year}{2019}).
\newblock \urlprefix\url{https://doi.org/10.1063/1.5088164}.
\newblock
  \eprint{https://pubs.aip.org/aip/apr/article-pdf/doi/10.1063/1.5088164/14577412/021314\_1\_online.pdf}.

\bibitem{Monroe2021}
\bibinfo{author}{Monroe, C.} \emph{et~al.}
\newblock \bibinfo{title}{Programmable quantum simulations of spin systems with
  trapped ions}.
\newblock \emph{\bibinfo{journal}{Rev. Mod. Phys.}}
  \textbf{\bibinfo{volume}{93}}, \bibinfo{pages}{025001}
  (\bibinfo{year}{2021}).
\newblock
  \urlprefix\url{https://link.aps.org/doi/10.1103/RevModPhys.93.025001}.

\bibitem{Knight2003quantum}
\bibinfo{author}{Knight, P.~L.} \emph{et~al.}
\newblock \bibinfo{title}{Quantum information processing with trapped ions}.
\newblock \emph{\bibinfo{journal}{Philosophical Transactions of the Royal
  Society of London. Series A: Mathematical, Physical and Engineering
  Sciences}} \textbf{\bibinfo{volume}{361}}, \bibinfo{pages}{1349--1361}
  (\bibinfo{year}{2003}).
\newblock
  \urlprefix\url{https://royalsocietypublishing.org/doi/abs/10.1098/rsta.2003.1205}.

\bibitem{blatt2008entangled}
\bibinfo{author}{Blatt, R.} \& \bibinfo{author}{Wineland, D.}
\newblock \bibinfo{title}{Entangled states of trapped atomic ions}.
\newblock \emph{\bibinfo{journal}{Nature}} \textbf{\bibinfo{volume}{453}},
  \bibinfo{pages}{1008--1015} (\bibinfo{year}{2008}).
\newblock \urlprefix\url{https://doi.org/10.1038/nature07125}.

\bibitem{Olmschenk2007}
\bibinfo{author}{Olmschenk, S.} \emph{et~al.}
\newblock \bibinfo{title}{Manipulation and detection of a trapped
  ${\mathrm{yb}}^{+}$ hyperfine qubit}.
\newblock \emph{\bibinfo{journal}{Phys. Rev. A}} \textbf{\bibinfo{volume}{76}},
  \bibinfo{pages}{052314} (\bibinfo{year}{2007}).
\newblock \urlprefix\url{https://link.aps.org/doi/10.1103/PhysRevA.76.052314}.

\bibitem{harty2014high-fidelity}
\bibinfo{author}{Harty, T.} \emph{et~al.}
\newblock \bibinfo{title}{High-fidelity preparation, gates, memory, and readout
  of a trapped-ion quantum bit}.
\newblock \emph{\bibinfo{journal}{Phys. Rev. Lett.}}
  \textbf{\bibinfo{volume}{113}} (\bibinfo{year}{2014}).
\newblock \urlprefix\url{http://dx.doi.org/10.1103/PhysRevLett.113.220501}.

\bibitem{Yang2022}
\bibinfo{author}{Yang, H.~X.} \emph{et~al.}
\newblock \bibinfo{title}{Realizing coherently convertible dual-type qubits
  with the same ion species}.
\newblock \emph{\bibinfo{journal}{Nature Physics}}
  \textbf{\bibinfo{volume}{18}}, \bibinfo{pages}{1058--1061}
  (\bibinfo{year}{2022}).
\newblock \urlprefix\url{https://doi.org/10.1038/s41567-022-01661-5}.

\bibitem{Allcock2021}
\bibinfo{author}{Allcock, D. T.~C.} \emph{et~al.}
\newblock \bibinfo{title}{{omg blueprint for trapped ion quantum computing with
  metastable states}}.
\newblock \emph{\bibinfo{journal}{Applied Physics Letters}}
  \textbf{\bibinfo{volume}{119}}, \bibinfo{pages}{214002}
  (\bibinfo{year}{2021}).
\newblock \urlprefix\url{https://doi.org/10.1063/5.0069544}.
\newblock
  \eprint{https://pubs.aip.org/aip/apl/article-pdf/doi/10.1063/5.0069544/13267942/214002\_1\_online.pdf}.

\bibitem{Kang2023}
\bibinfo{author}{Kang, M.}, \bibinfo{author}{Campbell, W.~C.} \&
  \bibinfo{author}{Brown, K.~R.}
\newblock \bibinfo{title}{Quantum error correction with metastable states of
  trapped ions using erasure conversion}.
\newblock \emph{\bibinfo{journal}{PRX Quantum}} \textbf{\bibinfo{volume}{4}},
  \bibinfo{pages}{020358} (\bibinfo{year}{2023}).
\newblock \urlprefix\url{https://link.aps.org/doi/10.1103/PRXQuantum.4.020358}.

\bibitem{Senko2015}
\bibinfo{author}{Senko, C.} \emph{et~al.}
\newblock \bibinfo{title}{Realization of a quantum integer-spin chain with
  controllable interactions}.
\newblock \emph{\bibinfo{journal}{Phys. Rev. X}} \textbf{\bibinfo{volume}{5}},
  \bibinfo{pages}{021026} (\bibinfo{year}{2015}).
\newblock \urlprefix\url{https://link.aps.org/doi/10.1103/PhysRevX.5.021026}.

\bibitem{Low2020}
\bibinfo{author}{Low, P.~J.}, \bibinfo{author}{White, B.~M.},
  \bibinfo{author}{Cox, A.~A.}, \bibinfo{author}{Day, M.~L.} \&
  \bibinfo{author}{Senko, C.}
\newblock \bibinfo{title}{Practical trapped-ion protocols for universal
  qudit-based quantum computing}.
\newblock \emph{\bibinfo{journal}{Phys. Rev. Res.}}
  \textbf{\bibinfo{volume}{2}}, \bibinfo{pages}{033128} (\bibinfo{year}{2020}).
\newblock
  \urlprefix\url{https://link.aps.org/doi/10.1103/PhysRevResearch.2.033128}.

\bibitem{low2023control}
\bibinfo{author}{Low, P.~J.}, \bibinfo{author}{White, B.} \&
  \bibinfo{author}{Senko, C.}
\newblock \bibinfo{title}{Control and readout of a 13-level trapped ion qudit}
  (\bibinfo{year}{2023}).
\newblock \eprint{2306.03340}.

\bibitem{Ringbauer2022}
\bibinfo{author}{Ringbauer, M.} \emph{et~al.}
\newblock \bibinfo{title}{A universal qudit quantum processor with trapped
  ions}.
\newblock \emph{\bibinfo{journal}{Nature Physics}}
  \textbf{\bibinfo{volume}{18}}, \bibinfo{pages}{1053--1057}
  (\bibinfo{year}{2022}).
\newblock \urlprefix\url{https://doi.org/10.1038/s41567-022-01658-0}.

\bibitem{Hrmo2023}
\bibinfo{author}{Hrmo, P.} \emph{et~al.}
\newblock \bibinfo{title}{Native qudit entanglement in a trapped ion quantum
  processor}.
\newblock \emph{\bibinfo{journal}{Nature Communications}}
  \textbf{\bibinfo{volume}{14}}, \bibinfo{pages}{2242} (\bibinfo{year}{2023}).
\newblock \urlprefix\url{https://doi.org/10.1038/s41467-023-37375-2}.

\bibitem{leibfried2003quantum}
\bibinfo{author}{Leibfried, D.}, \bibinfo{author}{Blatt, R.},
  \bibinfo{author}{Monroe, C.} \& \bibinfo{author}{Wineland, D.}
\newblock \bibinfo{title}{Quantum dynamics of single trapped ions}.
\newblock \emph{\bibinfo{journal}{Rev. Mod. Phys.}}
  \textbf{\bibinfo{volume}{75}}, \bibinfo{pages}{281--324}
  (\bibinfo{year}{2003}).

\bibitem{PhysRevLett.117.060505}
\bibinfo{author}{Gaebler, J.~P.} \emph{et~al.}
\newblock \bibinfo{title}{High-fidelity universal gate set for
  ${^{9}\mathrm{Be}}^{+}$ ion qubits}.
\newblock \emph{\bibinfo{journal}{Phys. Rev. Lett.}}
  \textbf{\bibinfo{volume}{117}}, \bibinfo{pages}{060505}
  (\bibinfo{year}{2016}).
\newblock
  \urlprefix\url{https://link.aps.org/doi/10.1103/PhysRevLett.117.060505}.

\bibitem{PhysRevLett.117.060504}
\bibinfo{author}{Ballance, C.~J.}, \bibinfo{author}{Harty, T.~P.},
  \bibinfo{author}{Linke, N.~M.}, \bibinfo{author}{Sepiol, M.~A.} \&
  \bibinfo{author}{Lucas, D.~M.}
\newblock \bibinfo{title}{High-fidelity quantum logic gates using trapped-ion
  hyperfine qubits}.
\newblock \emph{\bibinfo{journal}{Phys. Rev. Lett.}}
  \textbf{\bibinfo{volume}{117}}, \bibinfo{pages}{060504}
  (\bibinfo{year}{2016}).
\newblock
  \urlprefix\url{https://link.aps.org/doi/10.1103/PhysRevLett.117.060504}.

\bibitem{clark2021high}
\bibinfo{author}{Clark, C.~R.} \emph{et~al.}
\newblock \bibinfo{title}{High-fidelity bell-state preparation with ca+ 40
  optical qubits}.
\newblock \emph{\bibinfo{journal}{Physical Review Letters}}
  \textbf{\bibinfo{volume}{127}}, \bibinfo{pages}{130505}
  (\bibinfo{year}{2021}).

\bibitem{srinivas2021}
\bibinfo{author}{Srinivas, R.} \emph{et~al.}
\newblock \bibinfo{title}{High-fidelity laser-free universal control of trapped
  ion qubits}.
\newblock \emph{\bibinfo{journal}{Nature}} \textbf{\bibinfo{volume}{597}},
  \bibinfo{pages}{209--213} (\bibinfo{year}{2021}).
\newblock \urlprefix\url{https://doi.org/10.1038/s41586-021-03809-4}.

\bibitem{sutherland2024}
\bibinfo{author}{Tyler~Sutherland, R.} \& \bibinfo{author}{Foss-Feig, M.}
\newblock \bibinfo{title}{Laser-free trapped ion entangling gates with aese:
  adiabatic elimination of spin-motion entanglement}.
\newblock \emph{\bibinfo{journal}{New Journal of Physics}}
  \textbf{\bibinfo{volume}{26}}, \bibinfo{pages}{013013}
  (\bibinfo{year}{2024}).
\newblock \urlprefix\url{https://dx.doi.org/10.1088/1367-2630/ad19f9}.

\bibitem{paul1990electromagnetic}
\bibinfo{author}{Paul, W.}
\newblock \bibinfo{title}{Electromagnetic traps for charged and neutral
  particles}.
\newblock \emph{\bibinfo{journal}{Rev. Mod. Phys.}}
  \textbf{\bibinfo{volume}{62}}, \bibinfo{pages}{531--540}
  (\bibinfo{year}{1990}).

\bibitem{dehmelt1967radiofrequency}
\bibinfo{author}{Dehmelt, H.}
\newblock \bibinfo{title}{Radiofrequency spectroscopy of stored ions i:
  Storage}.
\newblock \emph{\bibinfo{journal}{Adv. At. Mol. Phys.}}
  \textbf{\bibinfo{volume}{3}}, \bibinfo{pages}{53--72} (\bibinfo{year}{1967}).

\bibitem{Brown1986Geonium}
\bibinfo{author}{Brown, L.~S.} \& \bibinfo{author}{Gabrielse, G.}
\newblock \bibinfo{title}{Geonium theory: Physics of a single electron or ion
  in a penning trap}.
\newblock \emph{\bibinfo{journal}{Rev. Mod. Phys.}}
  \textbf{\bibinfo{volume}{58}}, \bibinfo{pages}{233--311}
  (\bibinfo{year}{1986}).
\newblock \urlprefix\url{https://link.aps.org/doi/10.1103/RevModPhys.58.233}.

\bibitem{britton2012engineered}
\bibinfo{author}{Britton, J.~W.} \emph{et~al.}
\newblock \bibinfo{title}{{Engineered two-dimensional Ising interactions in a
  trapped-ion quantum simulator with hundreds of spins}}.
\newblock \emph{\bibinfo{journal}{Nature}} \textbf{\bibinfo{volume}{484}},
  \bibinfo{pages}{489--492} (\bibinfo{year}{2012}).

\bibitem{Bohnet2016}
\bibinfo{author}{Bohnet, J.~G.} \emph{et~al.}
\newblock \bibinfo{title}{Quantum spin dynamics and entanglement generation
  with hundreds of trapped ions}.
\newblock \emph{\bibinfo{journal}{Science}} \textbf{\bibinfo{volume}{352}},
  \bibinfo{pages}{1297--1301} (\bibinfo{year}{2016}).
\newblock
  \urlprefix\url{https://www.science.org/doi/abs/10.1126/science.aad9958}.
\newblock \eprint{https://www.science.org/doi/pdf/10.1126/science.aad9958}.

\bibitem{Gilmore2021}
\bibinfo{author}{Gilmore, K.~A.} \emph{et~al.}
\newblock \bibinfo{title}{Quantum-enhanced sensing of displacements and
  electric fields with two-dimensional trapped-ion crystals}.
\newblock \emph{\bibinfo{journal}{Science}} \textbf{\bibinfo{volume}{373}},
  \bibinfo{pages}{673--678} (\bibinfo{year}{2021}).
\newblock
  \urlprefix\url{https://www.science.org/doi/abs/10.1126/science.abi5226}.
\newblock \eprint{https://www.science.org/doi/pdf/10.1126/science.abi5226}.

\bibitem{Thompson2016}
\bibinfo{author}{Thompson, R.~C.}
\newblock \emph{\bibinfo{title}{PENNING TRAPS}}, chap.
  \bibinfo{chapter}{CHAPTER ONE}, \bibinfo{pages}{1--33}
  (\bibinfo{year}{2016}).
\newblock
  \urlprefix\url{https://www.worldscientific.com/doi/abs/10.1142/9781786340139_0001}.
\newblock
  \eprint{https://www.worldscientific.com/doi/pdf/10.1142/9781786340139_0001}.

\bibitem{Richerme2016}
\bibinfo{author}{Richerme, P.}
\newblock \bibinfo{title}{Two-dimensional ion crystals in radio-frequency traps
  for quantum simulation}.
\newblock \emph{\bibinfo{journal}{Phys. Rev. A}} \textbf{\bibinfo{volume}{94}},
  \bibinfo{pages}{032320} (\bibinfo{year}{2016}).
\newblock \urlprefix\url{https://link.aps.org/doi/10.1103/PhysRevA.94.032320}.

\bibitem{DOnofrio2021}
\bibinfo{author}{D'Onofrio, M.} \emph{et~al.}
\newblock \bibinfo{title}{Radial two-dimensional ion crystals in a linear paul
  trap}.
\newblock \emph{\bibinfo{journal}{Phys. Rev. Lett.}}
  \textbf{\bibinfo{volume}{127}}, \bibinfo{pages}{020503}
  (\bibinfo{year}{2021}).
\newblock
  \urlprefix\url{https://link.aps.org/doi/10.1103/PhysRevLett.127.020503}.

\bibitem{Wang2020}
\bibinfo{author}{Wang, Y.} \emph{et~al.}
\newblock \bibinfo{title}{Coherently manipulated 2d ion crystal in a monolithic
  paul trap}.
\newblock \emph{\bibinfo{journal}{Advanced Quantum Technologies}}
  \textbf{\bibinfo{volume}{3}}, \bibinfo{pages}{2000068}
  (\bibinfo{year}{2020}).
\newblock
  \urlprefix\url{https://onlinelibrary.wiley.com/doi/abs/10.1002/qute.202000068}.
\newblock
  \eprint{https://onlinelibrary.wiley.com/doi/pdf/10.1002/qute.202000068}.

\bibitem{Kiesenhofer2023}
\bibinfo{author}{Kiesenhofer, D.} \emph{et~al.}
\newblock \bibinfo{title}{Controlling two-dimensional coulomb crystals of more
  than 100 ions in a monolithic radio-frequency trap}.
\newblock \emph{\bibinfo{journal}{PRX Quantum}} \textbf{\bibinfo{volume}{4}},
  \bibinfo{pages}{020317} (\bibinfo{year}{2023}).
\newblock \urlprefix\url{https://link.aps.org/doi/10.1103/PRXQuantum.4.020317}.

\bibitem{Qiao2024tunable}
\bibinfo{author}{Qiao, M.} \emph{et~al.}
\newblock \bibinfo{title}{Tunable quantum simulation of spin models with a
  two-dimensional ion crystal}.
\newblock \emph{\bibinfo{journal}{Nature Physics}}  (\bibinfo{year}{2024}).
\newblock \urlprefix\url{https://doi.org/10.1038/s41567-023-02378-9}.

\bibitem{guo2023siteresolved}
\bibinfo{author}{Guo, S.~A.} \emph{et~al.}
\newblock \bibinfo{title}{A site-resolved 2d quantum simulator with hundreds of
  trapped ions under tunable couplings} (\bibinfo{year}{2023}).
\newblock \eprint{2311.17163}.

\bibitem{doi:10.1142/9789814529549}
\bibinfo{author}{HEUVELL, H. B. V. L. V.~D.}, \bibinfo{author}{WALRAVEN, J.
  T.~M.} \& \bibinfo{author}{REYNOLDS, M.~W.}
\newblock \emph{\bibinfo{title}{ATOMIC PHYSICS 15}}, \bibinfo{pages}{1--474}.
\newblock
  \urlprefix\url{https://www.worldscientific.com/doi/abs/10.1142/9789814529549}.
\newblock
  \eprint{https://www.worldscientific.com/doi/pdf/10.1142/9789814529549}.

\bibitem{wineland1998experimental}
\bibinfo{author}{Wineland, D.~J.} \emph{et~al.}
\newblock \bibinfo{title}{Experimental issues in coherent quantum-state
  manipulation of trapped atomic ions}.
\newblock \emph{\bibinfo{journal}{Journal of research of the National Institute
  of Standards and Technology}} \textbf{\bibinfo{volume}{103}},
  \bibinfo{pages}{259} (\bibinfo{year}{1998}).

\bibitem{kielpinski_QCCD}
\bibinfo{author}{Kielpinski, D.}, \bibinfo{author}{Monroe, C.} \&
  \bibinfo{author}{Wineland, D.~J.}
\newblock \bibinfo{title}{Architecture for a large-scale ion-trap quantum
  computer}.
\newblock \emph{\bibinfo{journal}{Nature}} \textbf{\bibinfo{volume}{417}},
  \bibinfo{pages}{709--711} (\bibinfo{year}{2002}).
\newblock \urlprefix\url{https://doi.org/10.1038/nature00784}.

\bibitem{chiaverini2005surface}
\bibinfo{author}{Chiaverini, J.} \emph{et~al.}
\newblock \bibinfo{title}{Surface-electrode architecture for ion-trap quantum
  information processing}.
\newblock \emph{\bibinfo{journal}{arXiv preprint quant-ph/0501147}}
  (\bibinfo{year}{2005}).

\bibitem{seidelin2006microfabricated}
\bibinfo{author}{Seidelin, S.} \emph{et~al.}
\newblock \bibinfo{title}{Microfabricated surface-electrode ion trap for
  scalable quantum information processing}.
\newblock \emph{\bibinfo{journal}{Physical review letters}}
  \textbf{\bibinfo{volume}{96}}, \bibinfo{pages}{253003}
  (\bibinfo{year}{2006}).

\bibitem{blakestad2010}
\bibinfo{author}{{Blakestad}, R.~B.}
\newblock \emph{\bibinfo{title}{{Transport of trapped-ion qubits within a
  scalable quantum processor}}}.
\newblock Ph.D. thesis, \bibinfo{school}{University of Colorado, Boulder}
  (\bibinfo{year}{2010}).

\bibitem{H1_data}
\bibinfo{author}{Baldwin, C.}
\newblock \bibinfo{title}{Quantinuum hardware specifications}
  \urlprefix\url{https://github.com/CQCL/quantinuum-hardware-specifications}.

\bibitem{PhysRevA.104.062440}
\bibinfo{author}{Gaebler, J.~P.} \emph{et~al.}
\newblock \bibinfo{title}{Suppression of midcircuit measurement crosstalk
  errors with micromotion}.
\newblock \emph{\bibinfo{journal}{Phys. Rev. A}}
  \textbf{\bibinfo{volume}{104}}, \bibinfo{pages}{062440}
  (\bibinfo{year}{2021}).
\newblock \urlprefix\url{https://link.aps.org/doi/10.1103/PhysRevA.104.062440}.

\bibitem{pino2021}
\bibinfo{author}{Pino, J.~M.} \emph{et~al.}
\newblock \bibinfo{title}{Demonstration of the trapped-ion quantum ccd computer
  architecture}.
\newblock \emph{\bibinfo{journal}{Nature}} \textbf{\bibinfo{volume}{592}},
  \bibinfo{pages}{209--213} (\bibinfo{year}{2021}).

\bibitem{moses2023}
\bibinfo{author}{Moses, S.~A.} \emph{et~al.}
\newblock \bibinfo{title}{A race-track trapped-ion quantum processor}.
\newblock \emph{\bibinfo{journal}{Physical Review X}}
  \textbf{\bibinfo{volume}{13}}, \bibinfo{pages}{041052}
  (\bibinfo{year}{2023}).

\bibitem{PhysRevA.86.032324}
\bibinfo{author}{Fowler, A.~G.}, \bibinfo{author}{Mariantoni, M.},
  \bibinfo{author}{Martinis, J.~M.} \& \bibinfo{author}{Cleland, A.~N.}
\newblock \bibinfo{title}{Surface codes: Towards practical large-scale quantum
  computation}.
\newblock \emph{\bibinfo{journal}{Phys. Rev. A}} \textbf{\bibinfo{volume}{86}},
  \bibinfo{pages}{032324} (\bibinfo{year}{2012}).
\newblock \urlprefix\url{https://link.aps.org/doi/10.1103/PhysRevA.86.032324}.

\bibitem{PhysRevLett.130.173202}
\bibinfo{author}{Burton, W.~C.} \emph{et~al.}
\newblock \bibinfo{title}{Transport of multispecies ion crystals through a
  junction in a radio-frequency paul trap}.
\newblock \emph{\bibinfo{journal}{Phys. Rev. Lett.}}
  \textbf{\bibinfo{volume}{130}}, \bibinfo{pages}{173202}
  (\bibinfo{year}{2023}).
\newblock
  \urlprefix\url{https://link.aps.org/doi/10.1103/PhysRevLett.130.173202}.

\bibitem{delaney2024scalable}
\bibinfo{author}{Delaney, R.~D.} \emph{et~al.}
\newblock \bibinfo{title}{Scalable multispecies ion transport in a grid based
  surface-electrode trap}.
\newblock \emph{\bibinfo{journal}{arXiv preprint arXiv:2403.00756}}
  (\bibinfo{year}{2024}).

\bibitem{Niffenegger2020}
\bibinfo{author}{Niffenegger, R.~J.} \emph{et~al.}
\newblock \bibinfo{title}{Integrated multi-wavelength control of an ion qubit}.
\newblock \emph{\bibinfo{journal}{Nature}} \textbf{\bibinfo{volume}{586}},
  \bibinfo{pages}{538--542} (\bibinfo{year}{2020}).
\newblock \urlprefix\url{https://doi.org/10.1038/s41586-020-2811-x}.

\bibitem{Mehta2020}
\bibinfo{author}{Mehta, K.~K.} \emph{et~al.}
\newblock \bibinfo{title}{Integrated optical multi-ion quantum logic}.
\newblock \emph{\bibinfo{journal}{Nature}} \textbf{\bibinfo{volume}{586}},
  \bibinfo{pages}{533--537} (\bibinfo{year}{2020}).
\newblock \urlprefix\url{https://doi.org/10.1038/s41586-020-2823-6}.

\bibitem{Ivory2021}
\bibinfo{author}{Ivory, M.} \emph{et~al.}
\newblock \bibinfo{title}{Integrated optical addressing of a trapped ytterbium
  ion}.
\newblock \emph{\bibinfo{journal}{Phys. Rev. X}} \textbf{\bibinfo{volume}{11}},
  \bibinfo{pages}{041033} (\bibinfo{year}{2021}).
\newblock \urlprefix\url{https://link.aps.org/doi/10.1103/PhysRevX.11.041033}.

\bibitem{kwon2023multi}
\bibinfo{author}{Kwon, J.} \emph{et~al.}
\newblock \bibinfo{title}{Multi-site integrated optical addressing of trapped
  ions}.
\newblock \emph{\bibinfo{journal}{arXiv preprint arXiv:2308.14918}}
  (\bibinfo{year}{2023}).

\bibitem{mordini2024multi}
\bibinfo{author}{Mordini, C.} \emph{et~al.}
\newblock \bibinfo{title}{Multi-zone trapped-ion qubit control in an integrated
  photonics qccd device}.
\newblock \emph{\bibinfo{journal}{arXiv preprint arXiv:2401.18056}}
  (\bibinfo{year}{2024}).

\bibitem{jain2024}
\bibinfo{author}{Jain, S.} \emph{et~al.}
\newblock \bibinfo{title}{Penning micro-trap for quantum computing}.
\newblock \emph{\bibinfo{journal}{Nature}}  (\bibinfo{year}{2024}).
\newblock \urlprefix\url{https://doi.org/10.1038/s41586-024-07111-x}.

\bibitem{Jain2020}
\bibinfo{author}{Jain, S.}, \bibinfo{author}{Alonso, J.},
  \bibinfo{author}{Grau, M.} \& \bibinfo{author}{Home, J.~P.}
\newblock \bibinfo{title}{Scalable arrays of micro-penning traps for quantum
  computing and simulation}.
\newblock \emph{\bibinfo{journal}{Phys. Rev. X}} \textbf{\bibinfo{volume}{10}},
  \bibinfo{pages}{031027} (\bibinfo{year}{2020}).
\newblock \urlprefix\url{https://link.aps.org/doi/10.1103/PhysRevX.10.031027}.

\bibitem{Duan2001}
\bibinfo{author}{Duan, L.~M.}, \bibinfo{author}{Lukin, M.~D.},
  \bibinfo{author}{Cirac, J.~I.} \& \bibinfo{author}{Zoller, P.}
\newblock \bibinfo{title}{Long-distance quantum communication with atomic
  ensembles and linear optics}.
\newblock \emph{\bibinfo{journal}{Nature}} \textbf{\bibinfo{volume}{414}},
  \bibinfo{pages}{413--418} (\bibinfo{year}{2001}).
\newblock \urlprefix\url{https://doi.org/10.1038/35106500}.

\bibitem{Monroe2013}
\bibinfo{author}{Monroe, C.} \& \bibinfo{author}{Kim, J.}
\newblock \bibinfo{title}{Scaling the ion trap quantum processor}.
\newblock \emph{\bibinfo{journal}{Science}} \textbf{\bibinfo{volume}{339}},
  \bibinfo{pages}{1164--1169} (\bibinfo{year}{2013}).

\bibitem{Moehring2007}
\bibinfo{author}{Moehring, D.~L.} \emph{et~al.}
\newblock \bibinfo{title}{Entanglement of single-atom quantum bits at a
  distance}.
\newblock \emph{\bibinfo{journal}{Nature}} \textbf{\bibinfo{volume}{449}},
  \bibinfo{pages}{68--71} (\bibinfo{year}{2007}).
\newblock \urlprefix\url{https://doi.org/10.1038/nature06118}.

\bibitem{Hucul2015}
\bibinfo{author}{Hucul, D.} \emph{et~al.}
\newblock \bibinfo{title}{Modular entanglement of atomic qubits using photons
  and phonons}.
\newblock \emph{\bibinfo{journal}{Nature Physics}}
  \textbf{\bibinfo{volume}{11}}, \bibinfo{pages}{37--42}
  (\bibinfo{year}{2015}).
\newblock \urlprefix\url{https://doi.org/10.1038/nphys3150}.

\bibitem{Stephenson2020}
\bibinfo{author}{Stephenson, L.~J.} \emph{et~al.}
\newblock \bibinfo{title}{High-rate, high-fidelity entanglement of qubits
  across an elementary quantum network}.
\newblock \emph{\bibinfo{journal}{Phys. Rev. Lett.}}
  \textbf{\bibinfo{volume}{124}}, \bibinfo{pages}{110501}
  (\bibinfo{year}{2020}).
\newblock
  \urlprefix\url{https://link.aps.org/doi/10.1103/PhysRevLett.124.110501}.

\bibitem{Carter2024}
\bibinfo{author}{Carter, A.~L.} \emph{et~al.}
\newblock \bibinfo{title}{{Ion trap with in-vacuum high numerical aperture
  imaging for a dual-species modular quantum computer}}.
\newblock \emph{\bibinfo{journal}{Review of Scientific Instruments}}
  \textbf{\bibinfo{volume}{95}}, \bibinfo{pages}{033201}
  (\bibinfo{year}{2024}).
\newblock \urlprefix\url{https://doi.org/10.1063/5.0180732}.
\newblock
  \eprint{https://pubs.aip.org/aip/rsi/article-pdf/doi/10.1063/5.0180732/19732485/033201\_1\_5.0180732.pdf}.

\bibitem{Schupp2021}
\bibinfo{author}{Schupp, J.} \emph{et~al.}
\newblock \bibinfo{title}{Interface between trapped-ion qubits and traveling
  photons with close-to-optimal efficiency}.
\newblock \emph{\bibinfo{journal}{PRX Quantum}} \textbf{\bibinfo{volume}{2}},
  \bibinfo{pages}{020331} (\bibinfo{year}{2021}).
\newblock \urlprefix\url{https://link.aps.org/doi/10.1103/PRXQuantum.2.020331}.

\bibitem{Molmer1999}
\bibinfo{author}{S\o{}rensen, A.} \& \bibinfo{author}{M\o{}lmer, K.}
\newblock \bibinfo{title}{Quantum computation with ions in thermal motion}.
\newblock \emph{\bibinfo{journal}{Phys. Rev. Lett.}}
  \textbf{\bibinfo{volume}{82}}, \bibinfo{pages}{1971--1974}
  (\bibinfo{year}{1999}).
\newblock \urlprefix\url{https://link.aps.org/doi/10.1103/PhysRevLett.82.1971}.

\bibitem{leibfried2003experimental}
\bibinfo{author}{Leibfried, D.} \emph{et~al.}
\newblock \bibinfo{title}{Experimental demonstration of a robust, high-fidelity
  geometric two ion-qubit phase gate}.
\newblock \emph{\bibinfo{journal}{Nature}} \textbf{\bibinfo{volume}{422}},
  \bibinfo{pages}{412--415} (\bibinfo{year}{2003}).
\newblock \urlprefix\url{http://dx.doi.org/10.1038/nature01492}.

\bibitem{schneider2012experimental}
\bibinfo{author}{Schneider, C.}, \bibinfo{author}{Porras, D.} \&
  \bibinfo{author}{Schaetz, T.}
\newblock \bibinfo{title}{{Experimental quantum simulations of many-body
  physics with trapped ions}}.
\newblock \emph{\bibinfo{journal}{Rep. Prog. Phys.}}
  \textbf{\bibinfo{volume}{75}} (\bibinfo{year}{2012}).

\bibitem{Pagano2020}
\bibinfo{author}{Pagano, G.} \emph{et~al.}
\newblock \bibinfo{title}{Quantum approximate optimization of the long-range
  ising model with a trapped-ion quantum simulator}.
\newblock \emph{\bibinfo{journal}{Proceedings of the National Academy of
  Sciences}} \textbf{\bibinfo{volume}{117}}, \bibinfo{pages}{25396--25401}
  (\bibinfo{year}{2020}).
\newblock \urlprefix\url{https://www.pnas.org/content/117/41/25396}.
\newblock \eprint{https://www.pnas.org/content/117/41/25396.full.pdf}.

\bibitem{Figgatt2019}
\bibinfo{author}{Figgatt, C.} \emph{et~al.}
\newblock \bibinfo{title}{Parallel entangling operations on a universal
  ion-trap quantum computer}.
\newblock \emph{\bibinfo{journal}{Nature}} \textbf{\bibinfo{volume}{572}},
  \bibinfo{pages}{368--372} (\bibinfo{year}{2019}).
\newblock \urlprefix\url{https://doi.org/10.1038/s41586-019-1427-5}.

\bibitem{Zhu2023}
\bibinfo{author}{Zhu, Y.} \emph{et~al.}
\newblock \bibinfo{title}{Pairwise-parallel entangling gates on orthogonal
  modes in a trapped-ion chain}.
\newblock \emph{\bibinfo{journal}{Advanced Quantum Technologies}}
  \textbf{\bibinfo{volume}{6}}, \bibinfo{pages}{2300056}
  (\bibinfo{year}{2023}).
\newblock
  \urlprefix\url{https://onlinelibrary.wiley.com/doi/abs/10.1002/qute.202300056}.
\newblock
  \eprint{https://onlinelibrary.wiley.com/doi/pdf/10.1002/qute.202300056}.

\bibitem{katz2022n}
\bibinfo{author}{Katz, O.}, \bibinfo{author}{Cetina, M.} \&
  \bibinfo{author}{Monroe, C.}
\newblock \bibinfo{title}{N-body interactions between trapped ion qubits via
  spin-dependent squeezing}.
\newblock \emph{\bibinfo{journal}{Physical Review Letters}}
  \textbf{\bibinfo{volume}{129}}, \bibinfo{pages}{063603}
  (\bibinfo{year}{2022}).

\bibitem{katz2023programmable}
\bibinfo{author}{Katz, O.}, \bibinfo{author}{Cetina, M.} \&
  \bibinfo{author}{Monroe, C.}
\newblock \bibinfo{title}{Programmable n-body interactions with trapped ions}.
\newblock \emph{\bibinfo{journal}{PRX Quantum}} \textbf{\bibinfo{volume}{4}},
  \bibinfo{pages}{030311} (\bibinfo{year}{2023}).

\bibitem{fang2023realization}
\bibinfo{author}{Fang, C.}, \bibinfo{author}{Wang, Y.}, \bibinfo{author}{Sun,
  K.} \& \bibinfo{author}{Kim, J.}
\newblock \bibinfo{title}{Realization of scalable cirac-zoller multi-qubit
  gates}.
\newblock \emph{\bibinfo{journal}{arXiv preprint arXiv:2301.07564}}
  (\bibinfo{year}{2023}).

\bibitem{katz2023demonstration}
\bibinfo{author}{Katz, O.}, \bibinfo{author}{Feng, L.},
  \bibinfo{author}{Risinger, A.}, \bibinfo{author}{Monroe, C.} \&
  \bibinfo{author}{Cetina, M.}
\newblock \bibinfo{title}{Demonstration of three-and four-body interactions
  between trapped-ion spins}.
\newblock \emph{\bibinfo{journal}{Nature Physics}}
  \textbf{\bibinfo{volume}{19}}, \bibinfo{pages}{1452--1458}
  (\bibinfo{year}{2023}).

\bibitem{tan2021domain}
\bibinfo{author}{Tan, W.~L.} \emph{et~al.}
\newblock \bibinfo{title}{Domain-wall confinement and dynamics in a quantum
  simulator}.
\newblock \emph{\bibinfo{journal}{Nature Physics}}
  \textbf{\bibinfo{volume}{17}}, \bibinfo{pages}{742--747}
  (\bibinfo{year}{2021}).
\newblock \urlprefix\url{https://doi.org/10.1038/s41567-021-01194-3}.

\bibitem{Davoudi2020towards}
\bibinfo{author}{Davoudi, Z.} \emph{et~al.}
\newblock \bibinfo{title}{Towards analog quantum simulations of lattice gauge
  theories with trapped ions}.
\newblock \emph{\bibinfo{journal}{Phys. Rev. Res.}}
  \textbf{\bibinfo{volume}{2}}, \bibinfo{pages}{023015} (\bibinfo{year}{2020}).
\newblock
  \urlprefix\url{https://link.aps.org/doi/10.1103/PhysRevResearch.2.023015}.

\bibitem{Nguyen2022digital}
\bibinfo{author}{Nguyen, N.~H.} \emph{et~al.}
\newblock \bibinfo{title}{Digital quantum simulation of the schwinger model and
  symmetry protection with trapped ions}.
\newblock \emph{\bibinfo{journal}{PRX Quantum}} \textbf{\bibinfo{volume}{3}},
  \bibinfo{pages}{020324} (\bibinfo{year}{2022}).
\newblock \urlprefix\url{https://link.aps.org/doi/10.1103/PRXQuantum.3.020324}.

\bibitem{Amitrano2023}
\bibinfo{author}{Amitrano, V.} \emph{et~al.}
\newblock \bibinfo{title}{Trapped-ion quantum simulation of collective neutrino
  oscillations}.
\newblock \emph{\bibinfo{journal}{Phys. Rev. D}}
  \textbf{\bibinfo{volume}{107}}, \bibinfo{pages}{023007}
  (\bibinfo{year}{2023}).
\newblock \urlprefix\url{https://link.aps.org/doi/10.1103/PhysRevD.107.023007}.

\bibitem{preskill2018simulating}
\bibinfo{author}{Preskill, J.}
\newblock \bibinfo{title}{Simulating quantum field theory with a quantum
  computer} (\bibinfo{year}{2018}).
\newblock \eprint{1811.10085}.

\bibitem{Banuls2020}
\bibinfo{author}{Ba{\~n}uls, M.~C.} \emph{et~al.}
\newblock \bibinfo{title}{Simulating lattice gauge theories within quantum
  technologies}.
\newblock \emph{\bibinfo{journal}{The European Physical Journal D}}
  \textbf{\bibinfo{volume}{74}}, \bibinfo{pages}{165} (\bibinfo{year}{2020}).
\newblock \urlprefix\url{https://doi.org/10.1140/epjd/e2020-100571-8}.

\bibitem{Bauer2023}
\bibinfo{author}{Bauer, C.~W.} \emph{et~al.}
\newblock \bibinfo{title}{Quantum simulation for high-energy physics}.
\newblock \emph{\bibinfo{journal}{PRX Quantum}} \textbf{\bibinfo{volume}{4}},
  \bibinfo{pages}{027001} (\bibinfo{year}{2023}).
\newblock \urlprefix\url{https://link.aps.org/doi/10.1103/PRXQuantum.4.027001}.

\bibitem{Bauer2023Quantum}
\bibinfo{author}{Bauer, C.~W.}, \bibinfo{author}{Davoudi, Z.},
  \bibinfo{author}{Klco, N.} \& \bibinfo{author}{Savage, M.~J.}
\newblock \bibinfo{title}{Quantum simulation of fundamental particles and
  forces}.
\newblock \emph{\bibinfo{journal}{Nature Reviews Physics}}
  \textbf{\bibinfo{volume}{5}}, \bibinfo{pages}{420--432}
  (\bibinfo{year}{2023}).
\newblock \urlprefix\url{https://doi.org/10.1038/s42254-023-00599-8}.

\bibitem{Gerritsma2010dirac}
\bibinfo{author}{Gerritsma, R.} \emph{et~al.}
\newblock \bibinfo{title}{Quantum simulation of the dirac equation}.
\newblock \emph{\bibinfo{journal}{Nature}} \textbf{\bibinfo{volume}{463}},
  \bibinfo{pages}{68--71} (\bibinfo{year}{2010}).
\newblock \urlprefix\url{https://doi.org/10.1038/nature08688}.

\bibitem{Wegner1971}
\bibinfo{author}{Wegner, F.~J.}
\newblock \bibinfo{title}{{Duality in Generalized Ising Models and Phase
  Transitions without Local Order Parameters}}.
\newblock \emph{\bibinfo{journal}{Journal of Mathematical Physics}}
  \textbf{\bibinfo{volume}{12}}, \bibinfo{pages}{2259--2272}
  (\bibinfo{year}{1971}).
\newblock \urlprefix\url{https://doi.org/10.1063/1.1665530}.
\newblock
  \eprint{https://pubs.aip.org/aip/jmp/article-pdf/12/10/2259/19106483/2259\_1\_online.pdf}.

\bibitem{Balian1975}
\bibinfo{author}{Balian, R.}, \bibinfo{author}{Drouffe, J.~M.} \&
  \bibinfo{author}{Itzykson, C.}
\newblock \bibinfo{title}{Gauge fields on a lattice. ii. gauge-invariant ising
  model}.
\newblock \emph{\bibinfo{journal}{Phys. Rev. D}} \textbf{\bibinfo{volume}{11}},
  \bibinfo{pages}{2098--2103} (\bibinfo{year}{1975}).
\newblock \urlprefix\url{https://link.aps.org/doi/10.1103/PhysRevD.11.2098}.

\bibitem{kormos2017real-time}
\bibinfo{author}{Kormos, M.}, \bibinfo{author}{Collura, M.},
  \bibinfo{author}{Tak{\'a}cs, G.} \& \bibinfo{author}{Calabrese, P.}
\newblock \bibinfo{title}{Real-time confinement following a quantum quench to a
  non-integrable model}.
\newblock \emph{\bibinfo{journal}{Nature Physics}}
  \textbf{\bibinfo{volume}{13}}, \bibinfo{pages}{246--249}
  (\bibinfo{year}{2017}).
\newblock \urlprefix\url{https://doi.org/10.1038/nphys3934}.

\bibitem{Verdel2020real}
\bibinfo{author}{Verdel, R.}, \bibinfo{author}{Liu, F.},
  \bibinfo{author}{Whitsitt, S.}, \bibinfo{author}{Gorshkov, A.~V.} \&
  \bibinfo{author}{Heyl, M.}
\newblock \bibinfo{title}{Real-time dynamics of string breaking in quantum spin
  chains}.
\newblock \emph{\bibinfo{journal}{Phys. Rev. B}}
  \textbf{\bibinfo{volume}{102}}, \bibinfo{pages}{014308}
  (\bibinfo{year}{2020}).
\newblock \urlprefix\url{https://link.aps.org/doi/10.1103/PhysRevB.102.014308}.

\bibitem{Mazza2019Suppression}
\bibinfo{author}{Mazza, P.~P.}, \bibinfo{author}{Perfetto, G.},
  \bibinfo{author}{Lerose, A.}, \bibinfo{author}{Collura, M.} \&
  \bibinfo{author}{Gambassi, A.}
\newblock \bibinfo{title}{Suppression of transport in nondisordered quantum
  spin chains due to confined excitations}.
\newblock \emph{\bibinfo{journal}{Phys. Rev. B}} \textbf{\bibinfo{volume}{99}},
  \bibinfo{pages}{180302} (\bibinfo{year}{2019}).
\newblock \urlprefix\url{https://link.aps.org/doi/10.1103/PhysRevB.99.180302}.

\bibitem{liu2019confined}
\bibinfo{author}{Liu, F.} \emph{et~al.}
\newblock \bibinfo{title}{Confined quasiparticle dynamics in long-range
  interacting quantum spin chains}.
\newblock \emph{\bibinfo{journal}{Phys. Rev. Lett.}}
  \textbf{\bibinfo{volume}{122}}, \bibinfo{pages}{150601}
  (\bibinfo{year}{2019}).
\newblock \urlprefix\url{https://link.aps.org/doi/10.1103/PhysRevA.99.043404}.

\bibitem{lerose2019quasilocalized}
\bibinfo{author}{Lerose, A.}, \bibinfo{author}{Zunkovic, B.},
  \bibinfo{author}{Silva, A.} \& \bibinfo{author}{Gambassi, A.}
\newblock \bibinfo{title}{Quasilocalized excitations induced by long-range
  interactions in translationally invariant quantum spin chains}.
\newblock \emph{\bibinfo{journal}{Phys. Rev. B}} \textbf{\bibinfo{volume}{99}},
  \bibinfo{pages}{121112} (\bibinfo{year}{2019}).
\newblock \urlprefix\url{https://link.aps.org/doi/10.1103/PhysRevB.99.121112}.

\bibitem{James2019Nonthermal}
\bibinfo{author}{James, A. J.~A.}, \bibinfo{author}{Konik, R.~M.} \&
  \bibinfo{author}{Robinson, N.~J.}
\newblock \bibinfo{title}{Nonthermal states arising from confinement in one and
  two dimensions}.
\newblock \emph{\bibinfo{journal}{Phys. Rev. Lett.}}
  \textbf{\bibinfo{volume}{122}}, \bibinfo{pages}{130603}
  (\bibinfo{year}{2019}).
\newblock
  \urlprefix\url{https://link.aps.org/doi/10.1103/PhysRevLett.122.130603}.

\bibitem{Kogut1975}
\bibinfo{author}{Kogut, J.} \& \bibinfo{author}{Susskind, L.}
\newblock \bibinfo{title}{Hamiltonian formulation of wilson's lattice gauge
  theories}.
\newblock \emph{\bibinfo{journal}{Phys. Rev. D}} \textbf{\bibinfo{volume}{11}},
  \bibinfo{pages}{395--408} (\bibinfo{year}{1975}).
\newblock \urlprefix\url{https://link.aps.org/doi/10.1103/PhysRevD.11.395}.

\bibitem{Muschik_2017}
\bibinfo{author}{Muschik, C.} \emph{et~al.}
\newblock \bibinfo{title}{U(1) wilson lattice gauge theories in digital quantum
  simulators}.
\newblock \emph{\bibinfo{journal}{New Journal of Physics}}
  \textbf{\bibinfo{volume}{19}}, \bibinfo{pages}{103020}
  (\bibinfo{year}{2017}).
\newblock \urlprefix\url{https://dx.doi.org/10.1088/1367-2630/aa89ab}.

\bibitem{Martinez2016real}
\bibinfo{author}{Martinez, E.~A.} \emph{et~al.}
\newblock \bibinfo{title}{Real-time dynamics of lattice gauge theories with a
  few-qubit quantum computer}.
\newblock \emph{\bibinfo{journal}{Nature}} \textbf{\bibinfo{volume}{534}},
  \bibinfo{pages}{516--519} (\bibinfo{year}{2016}).
\newblock \urlprefix\url{https://doi.org/10.1038/nature18318}.

\bibitem{Kokail2019}
\bibinfo{author}{Kokail, C.} \emph{et~al.}
\newblock \bibinfo{title}{Self-verifying variational quantum simulation of
  lattice models}.
\newblock \emph{\bibinfo{journal}{Nature}} \textbf{\bibinfo{volume}{569}},
  \bibinfo{pages}{355--360} (\bibinfo{year}{2019}).
\newblock \urlprefix\url{https://doi.org/10.1038/s41586-019-1177-4}.

\bibitem{Zache2019}
\bibinfo{author}{Zache, T.~V.} \emph{et~al.}
\newblock \bibinfo{title}{Dynamical topological transitions in the massive
  schwinger model with a $\ensuremath{\theta}$ term}.
\newblock \emph{\bibinfo{journal}{Phys. Rev. Lett.}}
  \textbf{\bibinfo{volume}{122}}, \bibinfo{pages}{050403}
  (\bibinfo{year}{2019}).
\newblock
  \urlprefix\url{https://link.aps.org/doi/10.1103/PhysRevLett.122.050403}.

\bibitem{jurcevic2017direct}
\bibinfo{author}{Jurcevic, P.} \emph{et~al.}
\newblock \bibinfo{title}{Direct observation of dynamical quantum phase
  transitions in an interacting many-body system}.
\newblock \emph{\bibinfo{journal}{Phys. Rev. Lett.}}
  \textbf{\bibinfo{volume}{119}}, \bibinfo{pages}{080501}
  (\bibinfo{year}{2017}).
\newblock
  \urlprefix\url{https://link.aps.org/doi/10.1103/PhysRevLett.119.080501}.

\bibitem{zhang2017observation}
\bibinfo{author}{Zhang, J.} \emph{et~al.}
\newblock \bibinfo{title}{Observation of a many-body dynamical phase transition
  with a 53-qubit quantum simulator}.
\newblock \emph{\bibinfo{journal}{Nature}} \textbf{\bibinfo{volume}{551}},
  \bibinfo{pages}{601--604} (\bibinfo{year}{2017}).
\newblock \urlprefix\url{http://dx.doi.org/10.1038/nature24654}.

\bibitem{heyl2018dynamical}
\bibinfo{author}{Heyl, M.}
\newblock \bibinfo{title}{Dynamical quantum phase transitions: a review}.
\newblock \emph{\bibinfo{journal}{Reports on Progress in Physics}}
  \textbf{\bibinfo{volume}{81}}, \bibinfo{pages}{054001}
  (\bibinfo{year}{2018}).
\newblock \urlprefix\url{https://doi.org/10.1088%2F1361-6633%2Faaaf9a}.

\bibitem{Mueller2023quantum}
\bibinfo{author}{Mueller, N.} \emph{et~al.}
\newblock \bibinfo{title}{Quantum computation of dynamical quantum phase
  transitions and entanglement tomography in a lattice gauge theory}.
\newblock \emph{\bibinfo{journal}{PRX Quantum}} \textbf{\bibinfo{volume}{4}},
  \bibinfo{pages}{030323} (\bibinfo{year}{2023}).
\newblock \urlprefix\url{https://link.aps.org/doi/10.1103/PRXQuantum.4.030323}.

\bibitem{Hauke2013}
\bibinfo{author}{Hauke, P.}, \bibinfo{author}{Marcos, D.},
  \bibinfo{author}{Dalmonte, M.} \& \bibinfo{author}{Zoller, P.}
\newblock \bibinfo{title}{Quantum simulation of a lattice schwinger model in a
  chain of trapped ions}.
\newblock \emph{\bibinfo{journal}{Phys. Rev. X}} \textbf{\bibinfo{volume}{3}},
  \bibinfo{pages}{041018} (\bibinfo{year}{2013}).
\newblock \urlprefix\url{https://link.aps.org/doi/10.1103/PhysRevX.3.041018}.

\bibitem{Andrade2022}
\bibinfo{author}{Andrade, B.} \emph{et~al.}
\newblock \bibinfo{title}{Engineering an effective three-spin hamiltonian in
  trapped-ion systems for applications in quantum simulation}.
\newblock \emph{\bibinfo{journal}{Quantum Science and Technology}}
  \textbf{\bibinfo{volume}{7}}, \bibinfo{pages}{034001} (\bibinfo{year}{2022}).
\newblock \urlprefix\url{https://doi.org/10.1088/2058-9565/ac5f5b}.

\bibitem{Davoudi2021towards}
\bibinfo{author}{Davoudi, Z.}, \bibinfo{author}{Linke, N.~M.} \&
  \bibinfo{author}{Pagano, G.}
\newblock \bibinfo{title}{Toward simulating quantum field theories with
  controlled phonon-ion dynamics: A hybrid analog-digital approach}.
\newblock \emph{\bibinfo{journal}{Phys. Rev. Res.}}
  \textbf{\bibinfo{volume}{3}}, \bibinfo{pages}{043072} (\bibinfo{year}{2021}).
\newblock
  \urlprefix\url{https://link.aps.org/doi/10.1103/PhysRevResearch.3.043072}.

\bibitem{meth2023simulating}
\bibinfo{author}{Meth, M.} \emph{et~al.}
\newblock \bibinfo{title}{Simulating 2d lattice gauge theories on a qudit
  quantum computer} (\bibinfo{year}{2023}).
\newblock \eprint{2310.12110}.

\bibitem{calajò2024digital}
\bibinfo{author}{Calajò, G.} \emph{et~al.}
\newblock \bibinfo{title}{Digital quantum simulation of a (1+1)d su(2) lattice
  gauge theory with ion qudits} (\bibinfo{year}{2024}).
\newblock \eprint{2402.07987}.

\bibitem{Pichler2016}
\bibinfo{author}{Pichler, T.}, \bibinfo{author}{Dalmonte, M.},
  \bibinfo{author}{Rico, E.}, \bibinfo{author}{Zoller, P.} \&
  \bibinfo{author}{Montangero, S.}
\newblock \bibinfo{title}{Real-time dynamics in u(1) lattice gauge theories
  with tensor networks}.
\newblock \emph{\bibinfo{journal}{Phys. Rev. X}} \textbf{\bibinfo{volume}{6}},
  \bibinfo{pages}{011023} (\bibinfo{year}{2016}).
\newblock \urlprefix\url{https://link.aps.org/doi/10.1103/PhysRevX.6.011023}.

\bibitem{Surace_2021}
\bibinfo{author}{Surace, F.~M.} \& \bibinfo{author}{Lerose, A.}
\newblock \bibinfo{title}{Scattering of mesons in quantum simulators}.
\newblock \emph{\bibinfo{journal}{New Journal of Physics}}
  \textbf{\bibinfo{volume}{23}}, \bibinfo{pages}{062001}
  (\bibinfo{year}{2021}).
\newblock \urlprefix\url{https://dx.doi.org/10.1088/1367-2630/abfc40}.

\bibitem{Milsted2022}
\bibinfo{author}{Milsted, A.}, \bibinfo{author}{Liu, J.},
  \bibinfo{author}{Preskill, J.} \& \bibinfo{author}{Vidal, G.}
\newblock \bibinfo{title}{Collisions of false-vacuum bubble walls in a quantum
  spin chain}.
\newblock \emph{\bibinfo{journal}{PRX Quantum}} \textbf{\bibinfo{volume}{3}},
  \bibinfo{pages}{020316} (\bibinfo{year}{2022}).
\newblock \urlprefix\url{https://link.aps.org/doi/10.1103/PRXQuantum.3.020316}.

\bibitem{bennewitz2024simulating}
\bibinfo{author}{Bennewitz, E.~R.} \emph{et~al.}
\newblock \bibinfo{title}{Simulating meson scattering on spin quantum
  simulators} (\bibinfo{year}{2024}).
\newblock \eprint{2403.07061}.

\bibitem{Gustafson2021}
\bibinfo{author}{Gustafson, E.}, \bibinfo{author}{Zhu, Y.},
  \bibinfo{author}{Dreher, P.}, \bibinfo{author}{Linke, N.~M.} \&
  \bibinfo{author}{Meurice, Y.}
\newblock \bibinfo{title}{Real-time quantum calculations of phase shifts using
  wave packet time delays}.
\newblock \emph{\bibinfo{journal}{Phys. Rev. D}}
  \textbf{\bibinfo{volume}{104}}, \bibinfo{pages}{054507}
  (\bibinfo{year}{2021}).
\newblock \urlprefix\url{https://link.aps.org/doi/10.1103/PhysRevD.104.054507}.

\bibitem{davoudi2024scattering}
\bibinfo{author}{Davoudi, Z.}, \bibinfo{author}{Hsieh, C.-C.} \&
  \bibinfo{author}{Kadam, S.~V.}
\newblock \bibinfo{title}{Scattering wave packets of hadrons in gauge theories:
  Preparation on a quantum computer} (\bibinfo{year}{2024}).
\newblock \eprint{2402.00840}.

\bibitem{Farrell2023}
\bibinfo{author}{Farrell, R.~C.} \emph{et~al.}
\newblock \bibinfo{title}{Preparations for quantum simulations of quantum
  chromodynamics in $1+1$ dimensions. ii. single-baryon
  $\ensuremath{\beta}$-decay in real time}.
\newblock \emph{\bibinfo{journal}{Phys. Rev. D}}
  \textbf{\bibinfo{volume}{107}}, \bibinfo{pages}{054513}
  (\bibinfo{year}{2023}).
\newblock \urlprefix\url{https://link.aps.org/doi/10.1103/PhysRevD.107.054513}.

\bibitem{Pantaleone1992}
\bibinfo{author}{Pantaleone, J.}
\newblock \bibinfo{title}{Neutrino oscillations at high densities}.
\newblock \emph{\bibinfo{journal}{Physics Letters B}}
  \textbf{\bibinfo{volume}{287}}, \bibinfo{pages}{128 -- 132}
  (\bibinfo{year}{1992}).
\newblock
  \urlprefix\url{http://www.sciencedirect.com/science/article/pii/037026939291887F}.

\bibitem{Cervia2019}
\bibinfo{author}{Cervia, M.~J.}, \bibinfo{author}{Patwardhan, A.~V.},
  \bibinfo{author}{Balantekin, A.~B.}, \bibinfo{author}{Coppersmith, S.~N.} \&
  \bibinfo{author}{Johnson, C.~W.}
\newblock \bibinfo{title}{Entanglement and collective flavor oscillations in a
  dense neutrino gas}.
\newblock \emph{\bibinfo{journal}{Phys. Rev. D}}
  \textbf{\bibinfo{volume}{100}}, \bibinfo{pages}{083001}
  (\bibinfo{year}{2019}).
\newblock \urlprefix\url{https://link.aps.org/doi/10.1103/PhysRevD.100.083001}.

\bibitem{Pehlivan2011}
\bibinfo{author}{Pehlivan, Y.}, \bibinfo{author}{Balantekin, A.~B.},
  \bibinfo{author}{Kajino, T.} \& \bibinfo{author}{Yoshida, T.}
\newblock \bibinfo{title}{Invariants of collective neutrino oscillations}.
\newblock \emph{\bibinfo{journal}{Phys. Rev. D}} \textbf{\bibinfo{volume}{84}},
  \bibinfo{pages}{065008} (\bibinfo{year}{2011}).
\newblock \urlprefix\url{https://link.aps.org/doi/10.1103/PhysRevD.84.065008}.

\bibitem{Landsman2019}
\bibinfo{author}{Landsman, K.~A.} \emph{et~al.}
\newblock \bibinfo{title}{Verified quantum information scrambling}.
\newblock \emph{\bibinfo{journal}{Nature}} \textbf{\bibinfo{volume}{567}},
  \bibinfo{pages}{61--65} (\bibinfo{year}{2019}).
\newblock \urlprefix\url{https://doi.org/10.1038/s41586-019-0952-6}.

\bibitem{Shapoval2023towardsquantum}
\bibinfo{author}{Shapoval, I.}, \bibinfo{author}{Su, V.~P.},
  \bibinfo{author}{Jong, W.~d.}, \bibinfo{author}{Urbanek, M.} \&
  \bibinfo{author}{Swingle, B.}
\newblock \bibinfo{title}{Towards {Q}uantum {G}ravity in the {L}ab on {Q}uantum
  {P}rocessors}.
\newblock \emph{\bibinfo{journal}{{Quantum}}} \textbf{\bibinfo{volume}{7}},
  \bibinfo{pages}{1138} (\bibinfo{year}{2023}).
\newblock \urlprefix\url{https://doi.org/10.22331/q-2023-10-12-1138}.

\bibitem{franke2023quantum}
\bibinfo{author}{Franke, J.} \emph{et~al.}
\newblock \bibinfo{title}{Quantum-enhanced sensing on optical transitions
  through finite-range interactions}.
\newblock \emph{\bibinfo{journal}{Nature}} \textbf{\bibinfo{volume}{621}},
  \bibinfo{pages}{740--745} (\bibinfo{year}{2023}).

\bibitem{giovannetti2011advances}
\bibinfo{author}{Giovannetti, V.}, \bibinfo{author}{Lloyd, S.} \&
  \bibinfo{author}{Maccone, L.}
\newblock \bibinfo{title}{Advances in quantum metrology}.
\newblock \emph{\bibinfo{journal}{Nature photonics}}
  \textbf{\bibinfo{volume}{5}}, \bibinfo{pages}{222--229}
  (\bibinfo{year}{2011}).

\bibitem{pezze2018quantum}
\bibinfo{author}{Pezze, L.}, \bibinfo{author}{Smerzi, A.},
  \bibinfo{author}{Oberthaler, M.~K.}, \bibinfo{author}{Schmied, R.} \&
  \bibinfo{author}{Treutlein, P.}
\newblock \bibinfo{title}{Quantum metrology with nonclassical states of atomic
  ensembles}.
\newblock \emph{\bibinfo{journal}{Reviews of Modern Physics}}
  \textbf{\bibinfo{volume}{90}}, \bibinfo{pages}{035005}
  (\bibinfo{year}{2018}).

\bibitem{monz201114}
\bibinfo{author}{Monz, T.} \emph{et~al.}
\newblock \bibinfo{title}{14-qubit entanglement: Creation and coherence}.
\newblock \emph{\bibinfo{journal}{Physical Review Letters}}
  \textbf{\bibinfo{volume}{106}}, \bibinfo{pages}{130506}
  (\bibinfo{year}{2011}).

\bibitem{pogorelov2021compact}
\bibinfo{author}{Pogorelov, I.} \emph{et~al.}
\newblock \bibinfo{title}{Compact ion-trap quantum computing demonstrator}.
\newblock \emph{\bibinfo{journal}{PRX Quantum}} \textbf{\bibinfo{volume}{2}},
  \bibinfo{pages}{020343} (\bibinfo{year}{2021}).

\bibitem{jeske2014quantum}
\bibinfo{author}{Jeske, J.}, \bibinfo{author}{Cole, J.~H.} \&
  \bibinfo{author}{Huelga, S.~F.}
\newblock \bibinfo{title}{Quantum metrology subject to spatially correlated
  markovian noise: restoring the heisenberg limit}.
\newblock \emph{\bibinfo{journal}{New Journal of Physics}}
  \textbf{\bibinfo{volume}{16}}, \bibinfo{pages}{073039}
  (\bibinfo{year}{2014}).

\bibitem{zhou2018achieving}
\bibinfo{author}{Zhou, S.}, \bibinfo{author}{Zhang, M.},
  \bibinfo{author}{Preskill, J.} \& \bibinfo{author}{Jiang, L.}
\newblock \bibinfo{title}{Achieving the heisenberg limit in quantum metrology
  using quantum error correction}.
\newblock \emph{\bibinfo{journal}{Nature communications}}
  \textbf{\bibinfo{volume}{9}}, \bibinfo{pages}{78} (\bibinfo{year}{2018}).

\bibitem{kitagawa1993squeezed}
\bibinfo{author}{Kitagawa, M.} \& \bibinfo{author}{Ueda, M.}
\newblock \bibinfo{title}{Squeezed spin states}.
\newblock \emph{\bibinfo{journal}{Physical Review A}}
  \textbf{\bibinfo{volume}{47}}, \bibinfo{pages}{5138} (\bibinfo{year}{1993}).

\bibitem{ma2011quantum}
\bibinfo{author}{Ma, J.}, \bibinfo{author}{Wang, X.}, \bibinfo{author}{Sun,
  C.-P.} \& \bibinfo{author}{Nori, F.}
\newblock \bibinfo{title}{Quantum spin squeezing}.
\newblock \emph{\bibinfo{journal}{Physics Reports}}
  \textbf{\bibinfo{volume}{509}}, \bibinfo{pages}{89--165}
  (\bibinfo{year}{2011}).

\bibitem{toth2009spin}
\bibinfo{author}{T{\'o}th, G.}, \bibinfo{author}{Knapp, C.},
  \bibinfo{author}{G{\"u}hne, O.} \& \bibinfo{author}{Briegel, H.~J.}
\newblock \bibinfo{title}{Spin squeezing and entanglement}.
\newblock \emph{\bibinfo{journal}{Physical Review A}}
  \textbf{\bibinfo{volume}{79}}, \bibinfo{pages}{042334}
  (\bibinfo{year}{2009}).

\bibitem{sinatra2022spin}
\bibinfo{author}{Sinatra, A.}
\newblock \bibinfo{title}{Spin-squeezed states for metrology}.
\newblock \emph{\bibinfo{journal}{Applied Physics Letters}}
  \textbf{\bibinfo{volume}{120}} (\bibinfo{year}{2022}).

\bibitem{leroux2010implementation}
\bibinfo{author}{Leroux, I.~D.}, \bibinfo{author}{Schleier-Smith, M.~H.} \&
  \bibinfo{author}{Vuleti{\'c}, V.}
\newblock \bibinfo{title}{Implementation of cavity squeezing of a collective
  atomic spin}.
\newblock \emph{\bibinfo{journal}{Physical Review Letters}}
  \textbf{\bibinfo{volume}{104}}, \bibinfo{pages}{073602}
  (\bibinfo{year}{2010}).

\bibitem{braverman2019near}
\bibinfo{author}{Braverman, B.} \emph{et~al.}
\newblock \bibinfo{title}{Near-unitary spin squeezing in yb 171}.
\newblock \emph{\bibinfo{journal}{Physical review letters}}
  \textbf{\bibinfo{volume}{122}}, \bibinfo{pages}{223203}
  (\bibinfo{year}{2019}).

\bibitem{combes2004states}
\bibinfo{author}{Combes, J.} \& \bibinfo{author}{Wiseman, H.}
\newblock \bibinfo{title}{States for phase estimation in quantum
  interferometry}.
\newblock \emph{\bibinfo{journal}{Journal of Optics B: Quantum and
  Semiclassical Optics}} \textbf{\bibinfo{volume}{7}}, \bibinfo{pages}{14}
  (\bibinfo{year}{2004}).

\bibitem{bohnet2016quantum}
\bibinfo{author}{Bohnet, J.~G.} \emph{et~al.}
\newblock \bibinfo{title}{Quantum spin dynamics and entanglement generation
  with hundreds of trapped ions}.
\newblock \emph{\bibinfo{journal}{Science}} \textbf{\bibinfo{volume}{352}},
  \bibinfo{pages}{1297--1301} (\bibinfo{year}{2016}).

\bibitem{crick2008two}
\bibinfo{author}{Crick, D.}, \bibinfo{author}{Ohadi, H.},
  \bibinfo{author}{Bhatti, I.}, \bibinfo{author}{Thompson, R.} \&
  \bibinfo{author}{Segal, D.}
\newblock \bibinfo{title}{Two-ion coulomb crystals of ca+ in a penning trap.}
\newblock \emph{\bibinfo{journal}{Optics Express}}
  \textbf{\bibinfo{volume}{16}}, \bibinfo{pages}{2351--2362}
  (\bibinfo{year}{2008}).

\bibitem{mavadia2013control}
\bibinfo{author}{Mavadia, S.} \emph{et~al.}
\newblock \bibinfo{title}{Control of the conformations of ion coulomb crystals
  in a penning trap}.
\newblock \emph{\bibinfo{journal}{Nature communications}}
  \textbf{\bibinfo{volume}{4}}, \bibinfo{pages}{2571} (\bibinfo{year}{2013}).

\bibitem{schleier2010squeezing}
\bibinfo{author}{Schleier-Smith, M.~H.}, \bibinfo{author}{Leroux, I.~D.} \&
  \bibinfo{author}{Vuleti{\'c}, V.}
\newblock \bibinfo{title}{Squeezing the collective spin of a dilute atomic
  ensemble by cavity feedback}.
\newblock \emph{\bibinfo{journal}{Physical Review A}}
  \textbf{\bibinfo{volume}{81}}, \bibinfo{pages}{021804}
  (\bibinfo{year}{2010}).

\bibitem{chen2014cavity}
\bibinfo{author}{Chen, Z.}, \bibinfo{author}{Bohnet, J.~G.},
  \bibinfo{author}{Weiner, J.~M.}, \bibinfo{author}{Cox, K.~C.} \&
  \bibinfo{author}{Thompson, J.~K.}
\newblock \bibinfo{title}{Cavity-aided nondemolition measurements for atom
  counting and spin squeezing}.
\newblock \emph{\bibinfo{journal}{Physical Review A}}
  \textbf{\bibinfo{volume}{89}}, \bibinfo{pages}{043837}
  (\bibinfo{year}{2014}).

\bibitem{comparin2022scalable}
\bibinfo{author}{Comparin, T.}, \bibinfo{author}{Mezzacapo, F.},
  \bibinfo{author}{Robert-de Saint-Vincent, M.} \& \bibinfo{author}{Roscilde,
  T.}
\newblock \bibinfo{title}{Scalable spin squeezing from spontaneous breaking of
  a continuous symmetry}.
\newblock \emph{\bibinfo{journal}{Physical Review Letters}}
  \textbf{\bibinfo{volume}{129}}, \bibinfo{pages}{113201}
  (\bibinfo{year}{2022}).

\bibitem{block2023universal}
\bibinfo{author}{Block, M.} \emph{et~al.}
\newblock \bibinfo{title}{A universal theory of spin squeezing}.
\newblock \emph{\bibinfo{journal}{arXiv preprint arXiv:2301.09636}}
  (\bibinfo{year}{2023}).

\bibitem{bornet2023scalable}
\bibinfo{author}{Bornet, G.} \emph{et~al.}
\newblock \bibinfo{title}{Scalable spin squeezing in a dipolar rydberg atom
  array}.
\newblock \emph{\bibinfo{journal}{Nature}} \textbf{\bibinfo{volume}{621}},
  \bibinfo{pages}{728--733} (\bibinfo{year}{2023}).

\bibitem{joshi2022observing}
\bibinfo{author}{Joshi, M.~K.} \emph{et~al.}
\newblock \bibinfo{title}{Observing emergent hydrodynamics in a long-range
  quantum magnet}.
\newblock \emph{\bibinfo{journal}{Science}} \textbf{\bibinfo{volume}{376}},
  \bibinfo{pages}{720--724} (\bibinfo{year}{2022}).

\bibitem{esposito2005emergence}
\bibinfo{author}{Esposito, M.} \& \bibinfo{author}{Gaspard, P.}
\newblock \bibinfo{title}{Emergence of diffusion in finite quantum systems}.
\newblock \emph{\bibinfo{journal}{Physical Review B}}
  \textbf{\bibinfo{volume}{71}}, \bibinfo{pages}{214302}
  (\bibinfo{year}{2005}).

\bibitem{castro2016emergent}
\bibinfo{author}{Castro-Alvaredo, O.~A.}, \bibinfo{author}{Doyon, B.} \&
  \bibinfo{author}{Yoshimura, T.}
\newblock \bibinfo{title}{Emergent hydrodynamics in integrable quantum systems
  out of equilibrium}.
\newblock \emph{\bibinfo{journal}{Physical Review X}}
  \textbf{\bibinfo{volume}{6}}, \bibinfo{pages}{041065} (\bibinfo{year}{2016}).

\bibitem{ye2020emergent}
\bibinfo{author}{Ye, B.}, \bibinfo{author}{Machado, F.},
  \bibinfo{author}{White, C.~D.}, \bibinfo{author}{Mong, R.~S.} \&
  \bibinfo{author}{Yao, N.~Y.}
\newblock \bibinfo{title}{Emergent hydrodynamics in nonequilibrium quantum
  systems}.
\newblock \emph{\bibinfo{journal}{Physical Review Letters}}
  \textbf{\bibinfo{volume}{125}}, \bibinfo{pages}{030601}
  (\bibinfo{year}{2020}).

\bibitem{sommer2011universal}
\bibinfo{author}{Sommer, A.}, \bibinfo{author}{Ku, M.}, \bibinfo{author}{Roati,
  G.} \& \bibinfo{author}{Zwierlein, M.~W.}
\newblock \bibinfo{title}{Universal spin transport in a strongly interacting
  fermi gas}.
\newblock \emph{\bibinfo{journal}{Nature}} \textbf{\bibinfo{volume}{472}},
  \bibinfo{pages}{201--204} (\bibinfo{year}{2011}).

\bibitem{moll2016evidence}
\bibinfo{author}{Moll, P.~J.}, \bibinfo{author}{Kushwaha, P.},
  \bibinfo{author}{Nandi, N.}, \bibinfo{author}{Schmidt, B.} \&
  \bibinfo{author}{Mackenzie, A.~P.}
\newblock \bibinfo{title}{Evidence for hydrodynamic electron flow in pdcoo2}.
\newblock \emph{\bibinfo{journal}{Science}} \textbf{\bibinfo{volume}{351}},
  \bibinfo{pages}{1061--1064} (\bibinfo{year}{2016}).

\bibitem{schuckert2020nonlocal}
\bibinfo{author}{Schuckert, A.}, \bibinfo{author}{Lovas, I.} \&
  \bibinfo{author}{Knap, M.}
\newblock \bibinfo{title}{Nonlocal emergent hydrodynamics in a long-range
  quantum spin system}.
\newblock \emph{\bibinfo{journal}{Physical Review B}}
  \textbf{\bibinfo{volume}{101}}, \bibinfo{pages}{020416}
  (\bibinfo{year}{2020}).

\bibitem{zu2021emergent}
\bibinfo{author}{Zu, C.} \emph{et~al.}
\newblock \bibinfo{title}{Emergent hydrodynamics in a strongly interacting
  dipolar spin ensemble}.
\newblock \emph{\bibinfo{journal}{Nature}} \textbf{\bibinfo{volume}{597}},
  \bibinfo{pages}{45--50} (\bibinfo{year}{2021}).

\bibitem{smith2016many}
\bibinfo{author}{Smith, J.} \emph{et~al.}
\newblock \bibinfo{title}{Many-body localization in a quantum simulator with
  programmable random disorder}.
\newblock \emph{\bibinfo{journal}{Nature Physics}}
  \textbf{\bibinfo{volume}{12}}, \bibinfo{pages}{907--911}
  (\bibinfo{year}{2016}).

\bibitem{morong2021observation}
\bibinfo{author}{Morong, W.} \emph{et~al.}
\newblock \bibinfo{title}{Observation of stark many-body localization without
  disorder}.
\newblock \emph{\bibinfo{journal}{Nature}} \textbf{\bibinfo{volume}{599}},
  \bibinfo{pages}{393--398} (\bibinfo{year}{2021}).

\bibitem{tantivasadakarn2023hierarchy}
\bibinfo{author}{Tantivasadakarn, N.}, \bibinfo{author}{Vishwanath, A.} \&
  \bibinfo{author}{Verresen, R.}
\newblock \bibinfo{title}{Hierarchy of topological order from finite-depth
  unitaries, measurement, and feedforward}.
\newblock \emph{\bibinfo{journal}{PRX Quantum}} \textbf{\bibinfo{volume}{4}},
  \bibinfo{pages}{020339} (\bibinfo{year}{2023}).

\bibitem{briegel2009measurement}
\bibinfo{author}{Briegel, H.~J.}, \bibinfo{author}{Browne, D.~E.},
  \bibinfo{author}{D{\"u}r, W.}, \bibinfo{author}{Raussendorf, R.} \&
  \bibinfo{author}{Van~den Nest, M.}
\newblock \bibinfo{title}{Measurement-based quantum computation}.
\newblock \emph{\bibinfo{journal}{Nature Physics}}
  \textbf{\bibinfo{volume}{5}}, \bibinfo{pages}{19--26} (\bibinfo{year}{2009}).

\bibitem{terhal2015quantum}
\bibinfo{author}{Terhal, B.~M.}
\newblock \bibinfo{title}{Quantum error correction for quantum memories}.
\newblock \emph{\bibinfo{journal}{Reviews of Modern Physics}}
  \textbf{\bibinfo{volume}{87}}, \bibinfo{pages}{307} (\bibinfo{year}{2015}).

\bibitem{potter2022entanglement}
\bibinfo{author}{Potter, A.~C.} \& \bibinfo{author}{Vasseur, R.}
\newblock \bibinfo{title}{Entanglement dynamics in hybrid quantum circuits}.
\newblock In \emph{\bibinfo{booktitle}{Entanglement in Spin Chains: From Theory
  to Quantum Technology Applications}}, \bibinfo{pages}{211--249}
  (\bibinfo{publisher}{Springer}, \bibinfo{year}{2022}).

\bibitem{fisher2023random}
\bibinfo{author}{Fisher, M.~P.}, \bibinfo{author}{Khemani, V.},
  \bibinfo{author}{Nahum, A.} \& \bibinfo{author}{Vijay, S.}
\newblock \bibinfo{title}{Random quantum circuits}.
\newblock \emph{\bibinfo{journal}{Annual Review of Condensed Matter Physics}}
  \textbf{\bibinfo{volume}{14}}, \bibinfo{pages}{335--379}
  (\bibinfo{year}{2023}).

\bibitem{Erhard2021}
\bibinfo{author}{Erhard, A.} \emph{et~al.}
\newblock \bibinfo{title}{Entangling logical qubits with lattice surgery}.
\newblock \emph{\bibinfo{journal}{Nature}} \textbf{\bibinfo{volume}{589}},
  \bibinfo{pages}{220--224} (\bibinfo{year}{2021}).
\newblock \urlprefix\url{https://doi.org/10.1038/s41586-020-03079-6}.

\bibitem{lieb1972finite}
\bibinfo{author}{Lieb, E.~H.} \& \bibinfo{author}{Robinson, D.~W.}
\newblock \bibinfo{title}{The finite group velocity of quantum spin systems}.
\newblock \emph{\bibinfo{journal}{Communications in mathematical physics}}
  \textbf{\bibinfo{volume}{28}}, \bibinfo{pages}{251--257}
  (\bibinfo{year}{1972}).

\bibitem{kitaev2010topological}
\bibinfo{author}{Kitaev, A.} \& \bibinfo{author}{Laumann, C.}
\newblock \bibinfo{title}{Topological phases and quantum computation}.
\newblock \emph{\bibinfo{journal}{Exact methods in low-dimensional statistical
  physics and quantum computing}} \bibinfo{pages}{101--125}
  (\bibinfo{year}{2010}).

\bibitem{iqbal2023topological}
\bibinfo{author}{Iqbal, M.} \emph{et~al.}
\newblock \bibinfo{title}{Topological order from measurements and feed-forward
  on a trapped ion quantum computer}.
\newblock \emph{\bibinfo{journal}{arXiv preprint arXiv:2302.01917}}
  (\bibinfo{year}{2023}).

\bibitem{foss2023experimental}
\bibinfo{author}{Foss-Feig, M.} \emph{et~al.}
\newblock \bibinfo{title}{Experimental demonstration of the advantage of
  adaptive quantum circuits}.
\newblock \emph{\bibinfo{journal}{arXiv preprint arXiv:2302.03029}}
  (\bibinfo{year}{2023}).

\bibitem{iqbal2024non}
\bibinfo{author}{Iqbal, M.} \emph{et~al.}
\newblock \bibinfo{title}{Non-abelian topological order and anyons on a
  trapped-ion processor}.
\newblock \emph{\bibinfo{journal}{Nature}} \textbf{\bibinfo{volume}{626}},
  \bibinfo{pages}{505--511} (\bibinfo{year}{2024}).

\bibitem{haghshenas2023probing}
\bibinfo{author}{Haghshenas, R.} \emph{et~al.}
\newblock \bibinfo{title}{Probing critical states of matter on a digital
  quantum computer}.
\newblock \emph{\bibinfo{journal}{arXiv preprint arXiv:2305.01650}}
  (\bibinfo{year}{2023}).

\bibitem{agrawal2023observing}
\bibinfo{author}{Agrawal, U.}, \bibinfo{author}{Lopez-Piqueres, J.},
  \bibinfo{author}{Vasseur, R.}, \bibinfo{author}{Gopalakrishnan, S.} \&
  \bibinfo{author}{Potter, A.~C.}
\newblock \bibinfo{title}{Observing quantum measurement collapse as a
  learnability phase transition}.
\newblock \emph{\bibinfo{journal}{arXiv preprint arXiv:2311.00058}}
  (\bibinfo{year}{2023}).

\bibitem{tantivasadakarn2021long}
\bibinfo{author}{Tantivasadakarn, N.}, \bibinfo{author}{Thorngren, R.},
  \bibinfo{author}{Vishwanath, A.} \& \bibinfo{author}{Verresen, R.}
\newblock \bibinfo{title}{Long-range entanglement from measuring
  symmetry-protected topological phases}.
\newblock \emph{\bibinfo{journal}{arXiv preprint arXiv:2112.01519}}
  (\bibinfo{year}{2021}).

\bibitem{levin2006detecting}
\bibinfo{author}{Levin, M.} \& \bibinfo{author}{Wen, X.-G.}
\newblock \bibinfo{title}{Detecting topological order in a ground state wave
  function}.
\newblock \emph{\bibinfo{journal}{Physical review letters}}
  \textbf{\bibinfo{volume}{96}}, \bibinfo{pages}{110405}
  (\bibinfo{year}{2006}).

\bibitem{kitaev2006topological}
\bibinfo{author}{Kitaev, A.} \& \bibinfo{author}{Preskill, J.}
\newblock \bibinfo{title}{Topological entanglement entropy}.
\newblock \emph{\bibinfo{journal}{Physical review letters}}
  \textbf{\bibinfo{volume}{96}}, \bibinfo{pages}{110404}
  (\bibinfo{year}{2006}).

\bibitem{bombin2010topological}
\bibinfo{author}{Bomb{\'\i}n, H.}
\newblock \bibinfo{title}{Topological order with a twist: Ising anyons from an
  abelian model}.
\newblock \emph{\bibinfo{journal}{Physical review letters}}
  \textbf{\bibinfo{volume}{105}}, \bibinfo{pages}{030403}
  (\bibinfo{year}{2010}).

\bibitem{nayak2008non}
\bibinfo{author}{Nayak, C.}, \bibinfo{author}{Simon, S.~H.},
  \bibinfo{author}{Stern, A.}, \bibinfo{author}{Freedman, M.} \&
  \bibinfo{author}{Sarma, S.~D.}
\newblock \bibinfo{title}{Non-abelian anyons and topological quantum
  computation}.
\newblock \emph{\bibinfo{journal}{Reviews of Modern Physics}}
  \textbf{\bibinfo{volume}{80}}, \bibinfo{pages}{1083} (\bibinfo{year}{2008}).

\bibitem{tantivasadakarn2023shortest}
\bibinfo{author}{Tantivasadakarn, N.}, \bibinfo{author}{Verresen, R.} \&
  \bibinfo{author}{Vishwanath, A.}
\newblock \bibinfo{title}{Shortest route to non-abelian topological order on a
  quantum processor}.
\newblock \emph{\bibinfo{journal}{Physical Review Letters}}
  \textbf{\bibinfo{volume}{131}}, \bibinfo{pages}{060405}
  (\bibinfo{year}{2023}).

\bibitem{propitius1995topological}
\bibinfo{author}{Propitius, M. d.~W.}
\newblock \bibinfo{title}{Topological interactions in broken gauge theories}.
\newblock \emph{\bibinfo{journal}{arXiv preprint hep-th/9511195}}
  (\bibinfo{year}{1995}).

\bibitem{bravyi2022adaptive}
\bibinfo{author}{Bravyi, S.}, \bibinfo{author}{Kim, I.},
  \bibinfo{author}{Kliesch, A.} \& \bibinfo{author}{Koenig, R.}
\newblock \bibinfo{title}{Adaptive constant-depth circuits for manipulating
  non-abelian anyons}.
\newblock \emph{\bibinfo{journal}{arXiv preprint arXiv:2205.01933}}
  (\bibinfo{year}{2022}).

\bibitem{mochon2004anyon}
\bibinfo{author}{Mochon, C.}
\newblock \bibinfo{title}{Anyon computers with smaller groups}.
\newblock \emph{\bibinfo{journal}{Physical Review A}}
  \textbf{\bibinfo{volume}{69}}, \bibinfo{pages}{032306}
  (\bibinfo{year}{2004}).

\bibitem{gottesman2013fault}
\bibinfo{author}{Gottesman, D.}
\newblock \bibinfo{title}{Fault-tolerant quantum computation with constant
  overhead}.
\newblock \emph{\bibinfo{journal}{arXiv preprint arXiv:1310.2984}}
  (\bibinfo{year}{2013}).

\bibitem{li2019measurement}
\bibinfo{author}{Li, Y.}, \bibinfo{author}{Chen, X.} \&
  \bibinfo{author}{Fisher, M.~P.}
\newblock \bibinfo{title}{Measurement-driven entanglement transition in hybrid
  quantum circuits}.
\newblock \emph{\bibinfo{journal}{Physical Review B}}
  \textbf{\bibinfo{volume}{100}}, \bibinfo{pages}{134306}
  (\bibinfo{year}{2019}).

\bibitem{noel2022measurement}
\bibinfo{author}{Noel, C.} \emph{et~al.}
\newblock \bibinfo{title}{Measurement-induced quantum phases realized in a
  trapped-ion quantum computer}.
\newblock \emph{\bibinfo{journal}{Nature Physics}}
  \textbf{\bibinfo{volume}{18}}, \bibinfo{pages}{760--764}
  (\bibinfo{year}{2022}).

\bibitem{schon2007sequential}
\bibinfo{author}{Sch{\"o}n, C.}, \bibinfo{author}{Hammerer, K.},
  \bibinfo{author}{Wolf, M.~M.}, \bibinfo{author}{Cirac, J.~I.} \&
  \bibinfo{author}{Solano, E.}
\newblock \bibinfo{title}{Sequential generation of matrix-product states in
  cavity qed}.
\newblock \emph{\bibinfo{journal}{Physical Review A}}
  \textbf{\bibinfo{volume}{75}}, \bibinfo{pages}{032311}
  (\bibinfo{year}{2007}).

\bibitem{kim2017holographic}
\bibinfo{author}{Kim, I.~H.}
\newblock \bibinfo{title}{Holographic quantum simulation}.
\newblock \emph{\bibinfo{journal}{arXiv preprint arXiv:1702.02093}}
  (\bibinfo{year}{2017}).

\bibitem{gopalakrishnan2019unitary}
\bibinfo{author}{Gopalakrishnan, S.} \& \bibinfo{author}{Lamacraft, A.}
\newblock \bibinfo{title}{Unitary circuits of finite depth and infinite width
  from quantum channels}.
\newblock \emph{\bibinfo{journal}{Physical Review B}}
  \textbf{\bibinfo{volume}{100}}, \bibinfo{pages}{064309}
  (\bibinfo{year}{2019}).

\bibitem{foss2022entanglement}
\bibinfo{author}{Foss-Feig, M.} \emph{et~al.}
\newblock \bibinfo{title}{Entanglement from tensor networks on a trapped-ion
  quantum computer}.
\newblock \emph{\bibinfo{journal}{Physical Review Letters}}
  \textbf{\bibinfo{volume}{128}}, \bibinfo{pages}{150504}
  (\bibinfo{year}{2022}).

\bibitem{foss2021holographic}
\bibinfo{author}{Foss-Feig, M.} \emph{et~al.}
\newblock \bibinfo{title}{Holographic quantum algorithms for simulating
  correlated spin systems}.
\newblock \emph{\bibinfo{journal}{Physical Review Research}}
  \textbf{\bibinfo{volume}{3}}, \bibinfo{pages}{033002} (\bibinfo{year}{2021}).

\bibitem{kim2017robust}
\bibinfo{author}{Kim, I.~H.} \& \bibinfo{author}{Swingle, B.}
\newblock \bibinfo{title}{Robust entanglement renormalization on a noisy
  quantum computer}.
\newblock \emph{\bibinfo{journal}{arXiv preprint arXiv:1711.07500}}
  (\bibinfo{year}{2017}).

\bibitem{liu2019variational}
\bibinfo{author}{Liu, J.-G.}, \bibinfo{author}{Zhang, Y.-H.},
  \bibinfo{author}{Wan, Y.} \& \bibinfo{author}{Wang, L.}
\newblock \bibinfo{title}{Variational quantum eigensolver with fewer qubits}.
\newblock \emph{\bibinfo{journal}{Physical Review Research}}
  \textbf{\bibinfo{volume}{1}}, \bibinfo{pages}{023025} (\bibinfo{year}{2019}).

\bibitem{PhysRevLett.128.010607}
\bibinfo{author}{Wei, Z.-Y.}, \bibinfo{author}{Malz, D.} \&
  \bibinfo{author}{Cirac, J.~I.}
\newblock \bibinfo{title}{Sequential generation of projected entangled-pair
  states}.
\newblock \emph{\bibinfo{journal}{Phys. Rev. Lett.}}
  \textbf{\bibinfo{volume}{128}}, \bibinfo{pages}{010607}
  (\bibinfo{year}{2022}).
\newblock
  \urlprefix\url{https://link.aps.org/doi/10.1103/PhysRevLett.128.010607}.

\bibitem{PhysRevLett.124.037201}
\bibinfo{author}{Zaletel, M.~P.} \& \bibinfo{author}{Pollmann, F.}
\newblock \bibinfo{title}{Isometric tensor network states in two dimensions}.
\newblock \emph{\bibinfo{journal}{Phys. Rev. Lett.}}
  \textbf{\bibinfo{volume}{124}}, \bibinfo{pages}{037201}
  (\bibinfo{year}{2020}).
\newblock
  \urlprefix\url{https://link.aps.org/doi/10.1103/PhysRevLett.124.037201}.

\bibitem{soejima2020isometric}
\bibinfo{author}{Soejima, T.} \emph{et~al.}
\newblock \bibinfo{title}{Isometric tensor network representation of string-net
  liquids}.
\newblock \emph{\bibinfo{journal}{Physical Review B}}
  \textbf{\bibinfo{volume}{101}}, \bibinfo{pages}{085117}
  (\bibinfo{year}{2020}).

\bibitem{PhysRevLett.101.110501}
\bibinfo{author}{Vidal, G.}
\newblock \bibinfo{title}{Class of quantum many-body states that can be
  efficiently simulated}.
\newblock \emph{\bibinfo{journal}{Phys. Rev. Lett.}}
  \textbf{\bibinfo{volume}{101}}, \bibinfo{pages}{110501}
  (\bibinfo{year}{2008}).
\newblock
  \urlprefix\url{https://link.aps.org/doi/10.1103/PhysRevLett.101.110501}.

\bibitem{anand2023holographic}
\bibinfo{author}{Anand, S.}, \bibinfo{author}{Hauschild, J.},
  \bibinfo{author}{Zhang, Y.}, \bibinfo{author}{Potter, A.~C.} \&
  \bibinfo{author}{Zaletel, M.~P.}
\newblock \bibinfo{title}{Holographic quantum simulation of entanglement
  renormalization circuits}.
\newblock \emph{\bibinfo{journal}{PRX Quantum}} \textbf{\bibinfo{volume}{4}},
  \bibinfo{pages}{030334} (\bibinfo{year}{2023}).

\bibitem{decross2023qubit}
\bibinfo{author}{DeCross, M.}, \bibinfo{author}{Chertkov, E.},
  \bibinfo{author}{Kohagen, M.} \& \bibinfo{author}{Foss-Feig, M.}
\newblock \bibinfo{title}{Qubit-reuse compilation with mid-circuit measurement
  and reset}.
\newblock \emph{\bibinfo{journal}{Physical Review X}}
  \textbf{\bibinfo{volume}{13}}, \bibinfo{pages}{041057}
  (\bibinfo{year}{2023}).

\bibitem{gray2021hyper}
\bibinfo{author}{Gray, J.} \& \bibinfo{author}{Kourtis, S.}
\newblock \bibinfo{title}{Hyper-optimized tensor network contraction}.
\newblock \emph{\bibinfo{journal}{Quantum}} \textbf{\bibinfo{volume}{5}},
  \bibinfo{pages}{410} (\bibinfo{year}{2021}).

\bibitem{napp2022efficient}
\bibinfo{author}{Napp, J.~C.}, \bibinfo{author}{La~Placa, R.~L.},
  \bibinfo{author}{Dalzell, A.~M.}, \bibinfo{author}{Brandao, F.~G.} \&
  \bibinfo{author}{Harrow, A.~W.}
\newblock \bibinfo{title}{Efficient classical simulation of random shallow 2d
  quantum circuits}.
\newblock \emph{\bibinfo{journal}{Physical Review X}}
  \textbf{\bibinfo{volume}{12}}, \bibinfo{pages}{021021}
  (\bibinfo{year}{2022}).

\bibitem{chertkov2022holographic}
\bibinfo{author}{Chertkov, E.} \emph{et~al.}
\newblock \bibinfo{title}{Holographic dynamics simulations with a trapped-ion
  quantum computer}.
\newblock \emph{\bibinfo{journal}{Nature Physics}}
  \textbf{\bibinfo{volume}{18}}, \bibinfo{pages}{1074--1079}
  (\bibinfo{year}{2022}).

\bibitem{chertkov2023characterizing}
\bibinfo{author}{Chertkov, E.} \emph{et~al.}
\newblock \bibinfo{title}{Characterizing a non-equilibrium phase transition on
  a quantum computer}.
\newblock \emph{\bibinfo{journal}{Nature Physics}}
  \textbf{\bibinfo{volume}{19}}, \bibinfo{pages}{1799--1804}
  (\bibinfo{year}{2023}).

\end{thebibliography}

\appendix

\end{document}